\documentclass[12pt]{article}
  \usepackage{ccaption}
  \captionnamefont{\bfseries}
  \captiontitlefont{\raggedright\doublespacing}
  \captiondelim{. }
\usepackage[normalem]{ulem}
\usepackage{xkeyval} %added for sky's compile problems
\usepackage{ctable}
\usepackage{url}
\usepackage{setspace}
\usepackage{natbib}

\usepackage{bm}
\bmdefine{\Bt}{t}
\bmdefine{\BX}{X}
\bmdefine{\BY}{Y}
\bmdefine{\BZ}{Z}
\bmdefine{\BB}{B}
\bmdefine{\BM}{M}
\bmdefine{\BD}{D}
\bmdefine{\Bi}{i}
\bmdefine{\Bj}{j}
\bmdefine{\Bx}{x}
\bmdefine{\By}{y}
\bmdefine{\Bz}{z}
\bmdefine{\Bw}{w}
\bmdefine{\Ba}{a}
\bmdefine{\Bb}{b}
\bmdefine{\Bc}{c}
\bmdefine{\Bh}{h}
\bmdefine{\Bg}{g}
\bmdefine{\Bu}{u}
\bmdefine{\Be}{e}

\newtheorem{thm}{Theorem}%[section]

\newtheorem{algorithm}[thm]{Algorithm}

\def\comment#1{\textit{[#1]}}
\def\comment#1{}

\def\jdlqed{\vbox{\hrule \hbox{\vrule\hbox to
5pt{\vbox to 6pt{\vfil}\hfil}\vrule}\hrule}}

\begin{document}

\doublespacing
% \thispagestyle{myheadings}
%\markright{HOST-SYMBIONT CODIVERGENCE}{}

\title{A Novel Test for Host-Symbiont Codivergence Indicates Ancient Origin of Fungal Endophytes in Grasses}
\author{C. L. Schardl$^1$, K. D. Craven$^{2}$, S. Speakman$^{3}$, A. Stromberg$^{3}$,\\  A. Lindstrom$^{3}$, and R. Yoshida$^{3}$\\ \\
$1${\it Department of Plant Pathology, 201F PSB, 1405 Veterans Drive,} \\{\it University of Kentucky, Lexington, KY 40546-0312.}\\ 
$2${\it Plant Biology Division, The Samuel Roberts Noble Foundation,} \\{\it 2510 Sam Noble Parkway, Ardmore, Oklahoma 73401.}\\ 
$3${\it Department of Statistics, University of Kentucky, Lexington, KY 40526-0027}\\
CLS, KDC, SS and RY contributed equally to this work}

%\date{}
% \thispagestyle{myheadings}
%\markright{HOST-SYMBIONT CODIVERGENCE}

%\maketitle
\noindent
HOST-SYMBIONT CODIVERGENCE
\vskip 0.2in 

\doublespacing

\begin{center}
{\Large \bf A Novel Test for Host-Symbiont Codivergence Indicates Ancient Origin of Fungal Endophytes in  Grasses}
\vskip 0.2in

{C. L. Schardl$^1$, K. D. Craven$^{2}$, S. Speakman$^{3}$, A. Stromberg$^{3}$,\\  A. Lindstrom$^{3}$, and R. Yoshida$^{3}$\\ 
$^1${\it Department of Plant Pathology, 201F PSB, 1405 Veterans Drive,} \\{\it University of Kentucky, Lexington, KY 40546-0312.}\\ 
$^2${\it Plant Biology Division, The Samuel Roberts Noble Foundation,} \\{\it 2510 Sam Noble Parkway, Ardmore, Oklahoma 73401.}\\ 
$^3${\it Department of Statistics, University of Kentucky, Lexington, KY 40526-0027}\\
CLS, KDC, SS and RY contributed equally to this work}
\end{center}

\noindent
Corresponding author: Ruriko Yoshida, \\
Department of Statistics, University of Kentucky, Lexington, KY 40526-0027\\
phone:(859) 257-5698, Fax:(859) 323-1973\\
email:\url{ruriko@ms.uky.edu},\\

\pagebreak

\raggedright

\hskip 0.5in {\it Abstract.--}
Significant phylogenetic codivergence between plant or animal hosts ($H$) and their symbionts or parasites ($P$)
 indicate the importance of their interactions on evolutionary time scales. However, valid and realistic methods to test 
for codivergence are not fully developed. One of the systems where possible codivergence has been of interest 
involves the large subfamily of temperate grasses (Pooideae) and their endophytic fungi (epichloae). These 
widespread symbioses often help protect host plants from herbivory and stresses, and affect species 
diversity and food web structures. Here we introduce the MRCALink (most-recent-common-ancestor link)
 method and use it to investigate the possibility of grass-epichlo\"e codivergence. MRCALink applied to 
ultrametric $H$ and $P$ trees identifies all corresponding nodes for pairwise comparisons of MRCA ages. 
The result is compared to the space of random $H$ and $P$ tree pairs estimated by a Monte Carlo method. 
Compared to tree reconciliation the method is less dependent on tree topologies (which often can be misleading), 
and it crucially improves on phylogeny-independent methods such as {\tt ParaFit} or the Mantel test by eliminating an extreme 
(but previously unrecognized) distortion of node-pair sampling. Analysis of 26 grass species-epichlo\"e species 
symbioses did not reject random association of $H$ and $P$ MRCA ages. However, when five obvious host 
jumps were removed the analysis significantly rejected random association and 
supported grass-endophyte codivergence. Interestingly, early cladogenesis events in the Pooideae 
corresponded to early cladogenesis events in epichloae, suggesting concomitant origins of this grass subfamily 
and its remarkable group of symbionts.  We also applied our method to the well-known gopher-louse data set. %from \citep{Hafner}.
\vskip 0.2in

\noindent
{\em key words}:  coevolution, grasses, endophytes, phylogenetic trees.

\pagebreak
%\section*{Introduction}

\hskip 0.5in 
Symbioses between cool-season grasses (Poaceae subfamily Pooideae) and fungi of genus {\em Epichlo\"e} (including 
their asexual derivatives in the genus {\em Neotyphodium}) are very widespread and occur in a broad taxonomic range of 
this important grass subfamily. These symbioses span the continuum from mutualistic to antagonistic interactions, 
making them an especially interesting model for evolution of mutualism, and particularly the possible role of 
codivergence \citep{Jackson,  Piano,  Schardl97,  Sullivan,  Tredway}. 
They have major ecological implications, affecting food web structures \citep{Omacini} and species diversity 
\citep{Clay99}. These endophytes extensively colonize host vegetative tissues without eliciting symptoms or 
defensive responses, and in many grass-epichlo\"e symbiota the endophyte colonizes the embryos and is vertically 
transmitted with exceptional efficiency \citep{Freeman,  Sampson33,  Sampson37}. 
The asexual endophytes rely upon vertical transmission for their propagation, whereas sexual ({\em Epichlo\"e}) 
species are also capable of horizontal transmission via meiotically derived ascospores \citep{Chung}. 
Mixed vertical and horizontal transmission strategies are also common. The tendency for vertical transmission is 
predicted to select for mutualism \citep{Bull,  Herre}. In fact, many vertically transmitted, and even some 
horizontally transmitted epichloae help protect their hosts from herbivores, nematodes, and other 
stressors \citep{Clay}. Grass-epichlo\"e symbioses occur in most tribes of the Pooideae, the highly speciose 
subfamily of temperate C3 grasses. Most but not all {\em Epichlo\"e} and {\em Neotyphodium} species are 
specialized to individual host species, genera or tribes in the Pooideae \citep{Schardl2005}. 
Therefore, it is reasonable to hypothesize a history of Pooideae-epichlo\"e coevolution extending back to the origin 
of this grass subfamily.  

\hskip 0.5in 
Previous analyses of the grass-epichlo\"e system have suggested some codivergence, along with some host species 
transfers (jumps) \citep{Jackson,  Schardl97}. However, methods for such comparative phylogenetic studies are 
in need of refinement. Interpreting evidence for codivergence based solely on congruence and reconciliation of 
tree topologies has significant shortcomings. Although tree reconciliation is useful particularly to assess strict 
cospeciation, precise mirror phylogenies for co-evolving hosts and symbionts can be an overly restrictive 
expectation \citep{Legendre, Page}.  For example, incomplete taxonomic sampling as well as lineage sorting of 
gene polymorphisms can result in topological incongruence between $H$ and $P$ trees, and deviations from 
strict cospeciation may mask tendencies for phylogenetic tracking. Refined methods are needed to assess 
significant patterns of codivergence without strict cospeciation. 
An attractive approach is to assess the correspondence in timing of cladogenesis events \citep{Hafner}. 
Methods that compare $H$ and $P$ pairwise distance matrices,  e.g., by the Mantel test \citep{Hafner90} or 
{\tt ParaFit} \citep{Legendre}, would seem to assess the timing of corresponding cladogenesis events directly. However, 
comprehensive pairwise distance methods suffer from grossly unequal sampling of corresponding cladogenesis events 
depending on the depth of the nodes representing those events in the actual $H$ and $P$ trees. The problem is illustrated by the simple examples in Figure 1 (and further documented in the Discussion). A statistical test for codivergence 
should evaluate the relationship between corresponding MRCA (most recent common ancestor) pairs without bias, 
but this is not equivalent to comparing matrices of all patristic distances between tree leaves (the extant sequences), 
simply because MRCAs deeper in the tree represent multiple descendant pairs, effectively weighting deeper (older) 
MRCAs over shallower MRCAs. 

\hskip 0.5in 
Here we introduce the MRCALink method to address the hypothesis that a group of hosts and symbionts (or parasites)
 have a significant degree of historical codivergence. The method is based on corresponding MRCA ages inferred from
 ultrametric maximum likelihood (ML) trees, but crucially samples each corresponding MRCA pair once and only once.
 From the set of sampled pairs of MRCA divergence times in corresponding $H$ and $P$ trees, we estimate the 
probability of similarity between the H and P trees by using randomly generated trees from the tree space. 
We apply this method to the grass-epichlo\"e system to assess evidence for codivergence and ancient origins of 
these symbioses.  This same method is also applied to a well-known gopher-louse data set for comparison purposes.

%\section*{Methods}
\vskip 0.2in

\begin{center}
{\textsc{Methods}}

\noindent{\it Fungal Endophyte Isolates and Endophyte-Infected Grasses}
\end{center}

\hskip 0.5in 
The cool-season grasses and their respective fungal endophytes are listed in Table \ref{table1}. All endophytes examined were from natural infections from which the corresponding natural host plant, or leaf material from this plant was available for chloroplast sequence analysis. Representatives of all available {\em Epichlo\"e} species and nonhybrid {\em Neotyphodium} species were included. In most cases where an {\em Epichlo\"e} species infects multiple host genera, isolates from each genus were sampled. The sole exception was {\em E. typhina}, for which codivergence can be rejected a priori due to its broad host range \citep{Leuchtmann, Craven}.
Many {\em Neotyphodium} species are interspecific hybrids, and therefore possess multiple genomes of distinct origin. These were usually excluded because of the difficulty in choosing the appropriate genome for analysis. However, four hybrid endophytes associated with {\em Lolium} species were included because they possess genomes in a clade (designated LAE, for {\em Lolium}-Associated Endophytes) that was unrepresented among those of known {\em Epichlo\"e} species. Each of these four species -- {\em Neotyphodium coenophialum}, {\em Neotyphodium occultans}, {\em Neotyphodium} sp. FaTG2 and {\em Neotyphodium} sp. FaTG3 -- had a distinct history of interspecific hybridization \citep{Tsai, Moon2000}.
\vskip 0.2in

\begin{center}
{{\it Sequencing of Chloroplast DNA (cpDNA) Non-Coding Regions}}
\end{center}

\hskip 0.5in 
Genomic DNA was isolated from 0.5-1.0 g of harvested endophyte-infected plant leaf material by the CTAB method \citep{Doyle}, and dissolved in 1 mL of purified water (Milli-Q; Millipore Corp., Bedford, Massachusetts).  DNA was quantified by bisbenzimide fluorescence, measured with a Hoefer (San Francisco, California) DyNA Quant 200 fluorometer.

\hskip 0.5in 
PCR amplification of one intron ({\em trnL} intron) and two intergenic spacers ({\em trnT}-{\em trnL}, {\em trnL}-{\em trnF}) from cpDNA was performed from total plant DNA with primers described by \cite{Taberlet}. Reactions were performed in 50 $\mu$L volumes containing 15 mM Tris-HCl, 1.5 mM MgCl$_2$, 50 mM KCl, pH 8.0 in the presence of 200 $\mu$M of each dNTP (Panvera, Madison, Wisconsin), 200 nM of each primer (Integrated DNA Technologies, Coralville, Iowa), 1.25 unit Amplitaq Gold DNA polymerase (Applied Biosystems, Foster City, California), and 10 ng of genomic DNA. Reactions were performed in a PE Applied Biosystems DNA thermal cycler, with a 9 min preheat step at 95$^o$C to activate the enzyme, followed by 35 cycles of 1 min at 95$^o$C, 1 min at 55$^o$C, and 1 min at 72$^o$C. All amplification products were verified by 0.8\% agarose gel electrophoresis, followed by visualization with ethidium bromide staining and UV fluorescence. The concentration of amplified products was estimated by comparison with a 100 bp quantitative ladder (Panvera). The amplified cpDNA products were purified with Qiaquick spin columns (Qiagen Inc., Valencia, California), then sequenced by the Sanger method with a BigDye Terminator Cycle version 1.0 or 3.1 sequencing kit (Applied Biosystems) or CEQ 2000 Dye Terminator Cycle Sequencing kit (Beckman-Coulter, Fullerton, California). The primers used in PCR were also used in sequencing, along with several primers designed for internal sequencing of amplified cpDNA fragments (Table \ref{table2}). Both DNA strands were sequenced. Products were separated by capillary electrophoresis on an Applied Biosystems 
model 310 genetic analyzer or on a CEQ 8000 genetic analyzer (Beckman-Coulter) at the University of Kentucky Advanced Genetic Technologies Center. Sequences were assembled with either Sequence Navigator software (Applied Biosystems) or Phrap (CodonCode Corporation, Dedham, Massachussets). Sequences were entered into GenBank as accession numbers AY450932--AY450949 and EU119353-EU119377.
\vskip 0.2in

\begin{center}
{{\it Endophyte DNA Sequences}}
\end{center}

\hskip 0.5in 
$\beta$-Tubulin and translation-elongation factor 1-$\alpha$ gene sequences for the endophytes included in this study were obtained previously \citep{Craven, Moon2004}. Employing a standardized gene nomenclature for {\em Epichlo\"e} and {\em Neotyphodium} species, these genes are designated {\em tubB} (formerly {\em tub}2) and {\em tefA} (formerly {\em tef}1), respectively.
\vskip 0.4cm

\begin{center}
{{\it Phylogenetic Tree Reconstruction and Analysis}}
\end{center}

\hskip 0.5in 
Sequences were aligned with the aid of {\tt PILEUP} implemented in {\tt SEQWeb} Version 1.1 with {\tt Wisconsin Package} Version 10 (Genetics Computer Group, Madison, Wisconsin). PILEUP parameters were adjusted empirically; a gap penalty of two and a gap extension penalty of zero resulted in reasonable alignment of intron-exon junctions and intron regions of endophyte sequences, and of intergenic spacer and intron regions of cpDNA sequences. Alignments were scrutinized and adjusted by eye, using tRNA or protein coding regions as anchor points. For phylogenetic analysis of the symbionts, sequences from {\em tubB} and {\em tefA} were concatenated to create a single, contiguous sequence of approximately 1400 bp for each endophyte, of which 357 bp was exon sequence and the remainder was intron sequence. For phylogenetic analysis of the hosts, sequences for both cpDNA intergenic regions ({\em trnT}-{\em trnL} and {\em trnL}-{\em trnF}) and the {\em trnL} intron were aligned individually then concatenated to give a combined alignment of approximately 2200 bp.

\hskip 0.5in 
Ultrametric trees were inferred with {\tt BEAST} v1.4.1 \citep{beast} with the general time reversible model with a proportion of invariable sites and a gamma distributed rate variation among sites (GTR+I+G). This model was selected by the software {\tt MrModelTest} vers. 2.0 (Johan A. A. Nylander, Uppsala University, \url{http://www.csit.fsu.edu/ nylander/mrmodeltest2readme.html})\citep{Posada} as the model of nucleotide substitution that best fits the data. Based on published phylogenetic inference for the grass subfamily Pooideae \citep{Soreng}, {\em Brachyelytrum erectum} was chosen as the outgroup for the grass phylogenies. The corresponding endophyte, {\em Epichlo\"e brachyelytri}, was the outgroup chosen for endophyte phylogenies. 
 Due to the lack of historical dates for interior nodes, {\tt BEAST} used a fixed 
substitution rate to reconstruct the trees.  This results in branch lengths
 (and tree height) being measured in substitutions per site.  The Markov chain Monte-Carlo
 method used by {\tt BEAST} was allowed to run for 5,000,000 steps.  Every 1000th step
 was recorded and analyzed for height, tree likelihood, and many other components.
  Preceding these recordings is a burn-in period equal to 10\% of the MCMC chain.  All data
from the burn-in period are discarded and the operators are not optimized during this time, thus preventing operators from optimizing incorrectly on trees that are still considered random at the beginning of each run. This process was done two independent runs from different tree topologies in order to avoid stacking at a local optimum, and resulted in a sample of 10,000 trees.

\hskip 0.5in 
The MRCALink algorithm reported herein requires ultrametric trees. However, for illustrative purposes only, 
%parametric trees 
phylograms were also inferred and posterior probabilities estimation with MrBayes version 3 \citep{Ronquist}, using a GTR+G model (lset nst = 6, rates = gamma). Four chains (three heated at temp = 0.2) were run for 450,000 generations, saving one out of every 100 trees (mcmc ngen = 450,000, printfreq = 10,000, samplefreq = 100, nchains = 4). The first 2000 trees (200,000 generations) were discarded as burn-in. This was an extremely conservative choice because likelihood values stabilized within 10,000 generations for both datasets. The 50\% majority rule consensus trees and posterior support values were determined from the 2500 trees sampled from the remaining 250,000 generations.
We ran each three independent runs with 200,000 iterations  
and always got the same consensus tree as shown in Figure 2 and 3.
%\ref{fig4} and \ref{figendo}.

\hskip 0.5in 
Sequence alignments and trees reconstructed via {\tt MrBayes} and {\tt BEAST} of plants and endophytes have been submitted to TreeBASE.
\vskip 0.2in

\begin{center}
{{\it The MRCALink Algorithm}}
\end{center}

\hskip 0.5in 
The MRCALink algorithm introduced here identifies and stores each corresponding $H$ and $P$ MRCA pair. 
Crucially, the data for each corresponding MRCA pair are selected only once for subsequent statistical analysis.  For example, if we have pairs of trees in Figure 1, then we pick MRCA pair $(7, 7')$
in the congruent tree four times from the set of all pairs of taxa in $H$ and $P$.
The MRCALink algorithm picks $(7, 7')$ only once instead of four times. Ultrametric  $H$ and $P$ trees must be used so 
that a unique age is estimated for each MRCA as half the patristic distance between any two of its descendant leaves. Trees must be strictly bifurcating for unique identification of valid $H$ and $P$ MRCA pairs. The {\tt BEAST} program outputs ultrametric and strictly bifurcating trees. Note that the method does not assume an equal number of taxa in $H$ and 
taxa in $P$, and also does not assume similar substitution rates in $H$ and $P$. Please see Appendix for the pseudo-code of the algorithm.

\hskip 0.5in 
Source codes for The MRCALink algorithm and the dissimilarity methods as well as data files are available at \url{http://www.ms.uky.edu/~ruriko/MRCALink/}.

\vskip 0.4cm

\begin{center}
{{\it Significance of Codivergence}}
\end{center}

\hskip 0.5in 
In this section we will discuss the two statistical methods used to test significance of codivergence
between the host and the parasite sets.  The first is a dissimilarity estimate between trees in the same treespace.
  The second test  uses {\tt ParaFit}  \citep{Legendre} to analyze the MRCA pairs sampled by the MRCALink algorithm.
The hypotheses are:
\begin{center}
Null hypothesis: Trees $T_H$ and $T_P$ are independent.

Alternative hypothesis: Trees $T_H$ and $T_P$ are not independent.
\end{center}

%\subsubsection*{Test via the dissimilarity method}

\hskip 0.5in {\it Test via the dissimilarity method.--}
We are interested in estimating the probability that the host and symbiont tree have some degree of dependence that may be due to a history of codivergence.  To this end, we use the sets of all pairwise differences
in $H$ and $P$ or the sets of pairwise differences 
in $H$ and $P$ from the MRCA pairs sampled by the MRCALink 
algorithm.  Let the sum of differences in uniquely estimated MRCA ages for 
trees $A$ and $B$ be $S(A,B)$.
The null hypothesis is that our $T_H$ and $T_P$ are
independent, so we generate a distribution of $S$ for pairs of 
unrelated random trees with the same number
of leaves and root-to-tip normalized distances (i.e., we normalize the heights 
of $T_H$ and $T_P$ to 1) as $T_H$ and $T_P$. 
Then we compare 
our $S(T_H, T_P)$ with this distribution.
If the p-value is significantly low ($< 0.05$), 
we reject the null hypothesis and conclude that there is evidence of codivergence between $T_H$ and $T_P$.
To calculate $S(A, B)$ with all pairwise distances, we take the 
sum of differences between pairwise distances for $A$ and $B$ over 
all pairwise distances.
To calculate $S(A, B)$ with the set of the MRCA pairs sampled by the 
MRCALink algorithm we take the sum of differences between pairwise 
distances for $A$ and $B$ over the set of the MRCA pairs sampled by the 
MRCALink algorithm.

\hskip 0.5in 
We generate $10,000$ random trees with the given 
branch lengths from the birth and death process (BDP) via {\tt evolver} from the 
{\tt PAML} package \citep{Yang} for each $T_H$ and $T_P$.  
For each tree, we used birth rate 0.5, death rate 0.5, and sampling fraction (SF = ratio of sample size to 
population size) 1, 0.5, 0.001, or 0.0005.  
We use the BDP model because it is biologically plausible \citep{Aris, Yang2} (also see more details on the Discussion section).

\hskip 0.5in 
Note the CPU time of this estimation with all pairwise distances is $O(Rn^2)$ 
and with the MRCALink algorithm it is $O(Rn^4)$, where $R$ is the number of
random trees and $n$ is the largest number of taxa in $H$ or $P$ 
(i.e., $n = \max\{n_1, n_2\}$ where  
$n_1$ is the number of taxa in $H$ and $n_2$ is the number of taxa in 
$P$). The CPU time of the dissimilarity method is a polynomial time 
algorithm in terms of the number of taxa.  Also note that this process
is easily distributed.  Thus, if we fix $R$ the computation time of
this method with all pairwise distances is $O(n^2)$ and the method with the MRCALink algorithm is $O(Rn^4)$.

%\subsubsection*{Testing via {\tt ParaFit}}

\hskip 0.5in {\it Testing via ParaFit.--}
The {\tt ParaFit} \citep{Legendre} method requires three input files: a relation table designating the hosts and their
parasites, and the coordinate representation of phylogenetic distances for $H$ and $P$.  These are the results
of PCA (Principle Component Analysis) on the respective distance matrices.  This method has been applied with
all  ${n_1 \choose 2}$ pairwise distances between two taxa in 
$H$ and  all ${n_2 \choose 2}$  pairwise distances between two taxa in $P$.
Instead, we use the distance matrices for $H$ and $P$ computed from the
set of MRCA pairs via the MRCALink algorithm.

\hskip 0.5in 
The following is an example of a distance matrix for all pairwise distances between any two taxa of a tree with 4 taxa.
  $d_{ij}$ represents a pairwise distance between taxa $i$ and $j$.
\[ \left( \begin{array}{cccc}
0 & d_{12} & d_{13} & d_{14} \\
d_{12} & 0 & d_{23} & d_{24} \\
d_{13} & d_{23} & 0 & d_{34} \\
d_{14} & d_{24} & d_{34} & 0 \\
 \end{array} \right)\] 

\hskip 0.5in 
Below is the same distance matrix using only the set of MRCA pairs provided by the MRCALink algorithm as used on the
congruent tree in Figure 1. Notice the removal of the distances between the sets of taxa (1,4), (2,3), and (2,4).
These represent the redundant information removed by the MRCALink algorithm.  The sets of taxa (1,3), (1,4), (2,3), and
(2,4) all have the same MRCA.  The distance between taxa 1 and 3 is sufficient and is the only one used in this 
reduced distance matrix.

\[ \left( \begin{array}{cccc}
0 & d_{12} & d_{13} & {\bf 0} \\
d_{12} & 0 & {\bf 0} & {\bf 0} \\
d_{13} & {\bf 0} & 0 & d_{34} \\
{\bf 0} & {\bf 0} & d_{34} & 0 \\
 \end{array} \right)\]

\hskip 0.5in 
Following is another distance matrix using only the set of MRCA pairs provided by the MRCALink algorithm as used on the
incongruent tree in Figure 1.  Notice the removal of the distances between the sets of taxa (2,3) and (2,4).
These represent the redundant information removed by the MRCALink algorithm.  The MRCALink algorithm has a smaller
effect on less congruent trees.

\[ \left( \begin{array}{cccc}
0 & d_{12} & d_{13} & d_{14} \\
d_{12} & 0 & {\bf 0} & {\bf 0} \\
d_{13} & {\bf 0} & 0 & d_{34} \\
d_{14} & {\bf 0} & d_{34} & 0 \\
 \end{array} \right)\]
\vskip 0.2in

\begin{center}
{{\it Random Tree generators for the CPU time test}}
\end{center}

\hskip 0.5in 
We generated random trees with 200 taxa by BDP with
0.5 birth rate, 0.5 death rate, and 0.0001 sampling fraction because \cite{Aris}, and \cite{Yang2} suggested that this is realistic model to generate random phylogenetic trees.  The heights of 
the trees are the same as $T_H$ and $T_P$ in the $T_4$ data set.  Since the computational time was 4-7 hours, we only took two sets of random trees for $T_H$ and $T_P$ and then we recorded the CPU time.  
\vskip 0.2in

\begin{center}
{\textsc{Results}}
%\hskip 0.6cm

{{\it Grass Phylogenies}}
%\hskip 0.6cm
\end{center}

\hskip 0.5in 
PCR amplification of {\em trnT}-{\em trnL} and {\em trnL}-{\em trnF} intergenic spacers and the {\em trnL} intron from endophyte-infected host plant genomic DNA yielded products of the sizes expected (approximately 850-950 bp, 400-450 bp for {\em trnT}-{\em trnL} and {\em trnL}-{\em trnF}, respectively; 350-600 bp for the {\em trnL} intron). The majority rule consensus tree with average branch lengths from MrBayes search on host cpDNA sequences is shown in Figure 2. In general, the inferred relationships among grass tribes were in good agreement with published grass phylogenies inferred by various genetic criteria \citep{Soreng, Kellogg, Catal}. Host grasses in the tribes Poeae, Agrostideae (syn = Aveneae), and Bromeae all formed monophyletic clades. Two of the three grasses in tribe Hordeeae (syn = Triticeae) -- {\em Hordeum brevisubulatum} and {\em Elymus canadensis} -- also grouped in a well-supported clade. However, {\em Hordelymus europaeus}, currently classified in the Hordeeae, grouped in the Bromeae clade basal to the {\em Bromus} species.

\hskip 0.5in 
Among the grasses in the tribe Poeae, a clear phylogenetic separation between the fine-leaf fescues ({\em Festuca} subg. {\em Festuca}) and the broad-leaf fescues ({\em Lolium} subg. {\em Schedonorus} = genus {\em Schedonorus}) was evident (Figure 2). Among the broad-leaf fescues in this analysis were three hexaploids previously classified as {\em Festuca arundinacea}, and representing plants from Europe (shown as {\em L. arundinaceum}), southern Spain ({\em Lolium} sp. P4074) and Algeria ({\em Lolium} sp. P4078) \citep{Tsai}. The latter two plants had closely related cpDNA sequences in a subclade basal to the European plant and species of {\em Lolium} subg. {\em Lolium}. Given this relationship, plants P4074 and P4078 are listed here as an undescribed {\em Lolium} species. The Poeae clade also had {\em Holcus mollis} in a basal position, which was the sister to the Agrostideae clade.

\hskip 0.5in 
The precise branching order of the most deeply rooted grasses was poorly resolved (Figure 2). The exception was {\em Brachypodium sylvaticum} (tribe Brachypodieae), which was placed nearest the clade comprising tribes Agrostideae, Poeae, Hordeeae and Bromeae. This result agreed with published studies \citep{Soreng, Kellogg}, which also indicate that tribe Brachyelytreae (represented by {\em Be. erectum})
diverged very early in the evolution of the cool-season grasses. Therefore, we chose {\em Be. erectum} to outgroup root the grass phylogeny.
\vskip 0.2in

\begin{center}
{{\it Endophyte Phylogenies}}
\end{center}

\hskip 0.5in 
The combined sequence data set ({\em tubB} + {\em tefA}) for the epichloae was approximately 1400 bp in length. The majority rule consensus tree from MrBayes search on the combined sequences (Figure 3) was in accord with the individual gene trees published earlier \citep{Moon2004}. Two large clades of the epichloae were well supported, and included only isolates from corresponding clades of the grasses. One of these included several species associated with either Agrostideae or its sister tribe Poeae, or both. This clade included {\em E. baconii, E. amarillans, E. festucae} and its closely related asexual species {\em N. lolii}, as well as an {\em Epichlo\"e} sp. isolate from {\em Holcus mollis}. Also included in this clade was the LAE ({\em Lolium}-associated endophyte) subclade of sequences from asexual symbionts of {\em Lolium} species. The endophytes with LAE genomes were all interspecific hybrids with additional genomes from {\em E. festucae}, {\em E. typhina} or {\em E. bromicola}; species that have contributed genomes to a large number of hybrid endophytes \citep{Moon2004}. Because the LAE genomes have not been identified in any sexual ({\em Epichlo\"e}) species, it is possible that the clade represents an old asexual lineage. If so, the LAE genome is very likely to have been transmitted vertically, because asexual epichloae are only known to transmit vertically in host maternal lineages  \citep{Chung, Brem}.

\hskip 0.5in 
Another clade grouped endophytes from sister grass tribes Bromeae and Hordeeae. This clade included {\em E. elymi, E. bromicola}, and asexual endophytes from {\em Bromus purgans, He. europaeus}, and {\em H. brevisubulatum}. 

\hskip 0.5in 
{\em Epichlo\"e glyceriae} and {\em Neotyphodium gansuense}, infecting grasses in tribes Meliceae and Stipeae respectively, were placed at basal positions relative to the other endophyte species. Another basal clade grouped {\em E. typhina} from {\em  L. perenne} with {\em E. sylvatica} and two asexual endophytes, {\em Neotyphodium typhinum} and {\em Neotyphodium aotearoae}. 
\vskip 0.4cm

\begin{center}
{{\it Comparison of Endophyte and Host Tree Topologies}}
\end{center}

\hskip 0.5in 
Several major groups or clades in the endophyte phylogeny corresponded to clades within the host phylogeny (Figure 4). For example, sister host tribes Agrostideae and Poeae mostly coincided with a similar grouping of their endophytes. Within tribe Poeae, a group containing {\em L. multiflorum, L. arundinaceum}, and {\em Lolium} sp. plants P4074 and P4078 was mirrored by the branching orders of their respective LAE-clade endophytes. The sister clade relationship of {\em Lolium} and {\em Festuca} species was reflected by the LAE and {\em E. festucae} sister clades, and the basal position of {\em Hol. mollis} within tribe Poeae was nearly matched by that of its corresponding symbiont. Similarly, endophytes of the sister tribes Bromeae and Hordeeae grouped in a clade. Grasses in basal host tribes Brachyelytreae, Stipeae, and Meliceae corresponded to basal endophyte clades {\em E. brachyelytri, N. gansuense}, and {\em E. glyceriae}, respectively. 

\hskip 0.5in 
Several instances of incongruence between host and endophyte phylogenies were also evident both within and across clades (Figure 4). Notable cases involved {\em E. typhina} and two related asexual endophytes, {\em N. typhinum} and {\em N. aotearoae}. All three of these endophytes infect grasses in tribes Poeae ({\em E. typhina} and {\em N. typhinum}) or Agrostideae ({\em N. aotearoae}), yet they grouped in a clade that was maximally divergent from the larger clade of endophytes from these grass tribes. Other examples of incongruence involved an {\em E. festucae} isolate from {\em Koeleria cristata} (tribe Agrostideae), {\em N. lolii} (an asexual derivative of {\em E. festucae}) from {\em  L. perenne}, and {\em E. elymi} from {\em Bro. purgans}.
\vskip 0.4cm

\begin{center}
{{\it Analyses of Codivergence}}
\end{center}

%\subsubsection*{Computation Results}

\hskip 0.5in {\it Computation results.--}
For each data set, we generated 10,000 ultrametric trees through {\tt BEAST} and chose the tree that had the maximum likelihood (Figure 4).  From this tree, we obtained corresponding pairs of $H$ (grasses) and $P$ (endophytes) MRCA ages (plotted in Figure 5). The significance of codivergence was estimated from randomly generated tree pairs with several sampling fractions (Table \ref{pvalues1}).  The lowest sampling fraction is probably the most biologically relevant \citep{Aris}.  We denote SF as a sampling fraction.  We first estimated p-values via the dissimilarity methods. For the full grass and endophyte trees at SF = 0.0005, we estimated $p = 0.123$ with the MRCALink algorithm and $p = 0.784$ with all pairwise distances. Thus, analysis of this data set did not reject the null hypothesis that the host and the parasite trees are independent. 

\hskip 0.5in 
An obvious source of discordance between endophyte and host trees was {\em E. typhina} and related endophytes. Both {\em N. lolii} and the {\em E. typhina} isolate in this study were from {\em L. perenne}, but the endophytes were maximally divergent from one another.
 Previous surveys have indicated that {\em E. typhina} has an unusually broad host range, and is ancestral to asexual endophytes (such as {\em N. typhinum}) in several other grasses (Leuchtmann and Schardl 1998; Moon et al. 2004; Gentile et al. 2005).
Therefore, the first trimmed tree set, $T_1$, had these taxa removed, and the host and endophyte $T_1$ trees were estimated. 
Analysis of the $T_1$ set gave lower p-values ($p < 0.001$ with the MRCALink algorithm and $p = 0.117$ with all pairwise distances) (Table \ref{pvalues1}).  Thus, the MRCALink analysis of this dataset supported dependence of the trees, although analysis with all pairwise distances did not. 
The hypothesis that the entire clade including {\em E. typhina}, {\em N. typhinum, E. sylvatica}, and {\em N. aotearoae} contributed most of the discordance was tested by eliminating all four of these taxa and their hosts in tree set $T_2$. The calculated p-values for $T_2$ ($p < 0.001$ with MRCALink and $p = 0.093$ with all pairwise distances) were comparable to those of $T_1$. 

\hskip 0.5in 
Trimmed set $T_3$ had {\em E. typhina} and {\em N. typhinum} and their hosts removed, as well as other taxa that appeared likely to represent jumps of endophytes between divergent hosts; namely {\em  N. lolii} and its host {\em L. perenne}, the {\em E. festucae-K. cristata} symbiotum, and the {\em E. elymi-Bro. purgans} symbiotum. 
The basis for considering these to be likely jumps was that the symbiont species were much more common on other grass genera: {\em E. festucae} on {\em Festuca} species, and {\em E. elymi} on {\em Elymus} species \citep{Craven, Moon2004}.
The p-values for the taxa in the $T_3$ set were $p < 0.001$ (Table \ref{pvalues1})  with the MRCALink algorithm and $p = 0.064$  with all pairwise distances (Table \ref{pvalues1}).  
Removal of all taxa that had been removed for $T_2$ and $T_3$ gave set $T_4$, for which we found significant evidence to reject the null hypothesis by both approaches.

\hskip 0.5in 
If codivergence has been an important trend since the origin of the epichloae and the pooid grasses, and if the DNA regions analyzed for these fungi and their hosts have had comparable substitution rates in the regions analyzed, then the estimated heights ($h$) of their respective phylogenetic trees should be comparable. Very different tree heights could be due to a lack of long-term codivergence of $H$ and $P$, or to a large difference in substitution rates for the regions chosen for analysis.
The tree heights estimated by  {\tt BEAST} in substitutions per site were as follows: for the full data set, $h(T_H)$ = 0.048623 and $h(T_P)$ = 0.028991; for $T_1$, $h(T_H)$ = 0.046534 and $h(T_P)$ = 0.028633; for $T_2$, $h(T_H)$ = 0.045813 and $h(T_P)$ = 0.028924; for $T_3$, $h(T_H)$ = 0.048321 and $h(T_P)$ = 0.028367; for $T_4$, $h(T_H)$ = 0.045188 and $h(T_P)$ = 0.027939.
Thus, for all of these data sets the inferred host tree heights were similar, and the inferred endophyte tree heights were similar. For both hosts and endophytes, almost all of the sequence analyzed was noncoding. The host data set was mainly intergenic sequence from chloroplast DNA, and most of the endophyte data comprised nuclear intronic sequence. The tree height estimates suggested that the substitution rate of the host sequences has been 1.58 to 1.70 times the substitution rate of the endophyte sequences. Thus, the estimated substitution rates were comparable, lending additional support to the hypothesis that the Pooideae and the epichloae originated at approximately the same time.

\hskip 0.5in 
We tested if the analyses reject sub-optimal parasite and host trees with no detectable host jumps.  
Instead of choosing the tree with maximum likelihood, we chose the three samples with likelihood values closest to the mean of likelihood values from the set of all trees sampled by {\tt BEAST}. 
For example, the maximum likelihood tree for the full plant data set that was used
in our methods had a negative log likelihood of -6771.2947.  The sub-optimal samples had negative log likelihoods of
-6783.4135, -6783.4134, and -6783.4076.  We then calculated p-values by the 
dissimilarity method (Table \ref{pvalues2}) and {\tt ParaFit} (Table 
\ref{pvalues4}). 

 \hskip 0.5in 
Application of the dissimilarity methods to the sub-optimal trees did not provide evidence to reject the null hypothesis (Table \ref{pvalues2}). 
The dissimilarity method with the MRCALink algorithm obtained significant p-values for some but not all samples.
%With the dissimilarity method with the MRCALink algorithm we obtained 
%significant p-values for some but not all samples. 

Computations were conducted on a Dual Core Pentium L2400 1.66 GHz PC machine in IBM Thinkpad laptop X60S with 2 GB RAM running Fedora Core 6 Linux.
For the dissimilarity method with all pairwise distances, 
calculating the p-value has the time complexity $O(n^2)$ where
$n$ is the largest number of taxa in $H$ or $P$ (i.e., 
$n = \max\{|H|, |P|\}$).  
Thus, this is a polynomial time algorithm in terms of the number of taxa.
Analyses of grass-endophyte data sets with the dissimilarity method with all pairwise distances
required CPU time of 39.5 sec for the full 26 taxon pairs, 33.481 sec for $T_1$, 27.2 sec for $T_2$, 25.0 sec for $T_3$, and 20.1 sec for $T_4$.  For computational time simulation analysis of our method with randomly generated 200-taxon trees with all pairwise distances, the dissimilarity method took 3.59 hr of CPU time.

\hskip 0.5in 
For the dissimilarity method with the MRCALink algorithm,
calculating the p-value has the time complexity $O(n^4)$ where
$n$ is the largest number of taxa in $H$ or $P$ (i.e., 
$n = \max\{|H|, |P|\}$).  
Thus, this is also a polynomial time algorithm in terms of the number of taxa.
Analyses of grass-endophyte data sets with the MRCALink algorithm
required CPU time of 2 min 49 sec for the 
full 26 taxon pairs, 2 min 16 sec for $T_1$, 1 min 41 sec for $T_2$,  
1 min 33 sec for $T_3$, and 1 min 9 sec for $T_4$.  
The CPU time of our method with 200 taxa was 7 hr. 

Note that to estimate the p-value using random trees, we can easily
distribute computations for both methods.  
Thus, this estimation method could be applied to host-parasite associations 
with several hundred taxa.

%\subsubsection*{Results with {\tt ParaFit}}

\hskip 0.5in {\it Results with {\tt ParaFit}.--}
We also analyzed the full grass and endophyte data sets $T_1$ through $T_4$ 
with {\tt ParaFit}, again using either all pairwise patristic distances or only those selected by MRCALink. For the former, we used {\tt ParaFit} with distance matrices for the data sets $H$ and $P$ after applying the PCA to their distance matrices.
 For the latter, we substituted 0 for some elements in order to represent only the set of MRCA pairs sampled by the MRCALink algorithm.  Note that {\tt ParaFit} takes the PCA, not distance matrices, so even if we remove some elements the resulting PCA differs.
Both approaches gave p-values all $< 0.001$, indicating rejection of the null hypothesis.  Thus, {\tt ParaFit} appeared to be highly sensitive to any dependence between trees.  

\hskip 0.5in 
Most p-values for  the sampled sub-optimal trees were significantly higher than 
p-values for the ML trees.  The p-value of sample 1 from the
$T_4$ data set was especially high compared to the p-value with the respective ML trees (Table \ref{pvalues4}).  
\vskip 0.4cm

\begin{center}
{{\it Results with Gopher-Louse Data Sets}}
\end{center}

\hskip 0.5in 
We also tested our methods with a well-known gopher-louse data set. 
The data set in \cite{Hafner} (full data set) contains 17 taxa of lice and 15 taxa of gophers, whereas \cite{huelsenbeck} have trimmed host-parasite pairs representing apparent host jumps: louse species {\em Geomydoecus thomomyus}, {\em Geomydoecus actuosi}, {\em Thomomydoecus barbarae} and 
{\em Thomomydoecus minor}, and gopher species {\em Thomomys talpoides} and {\em Thomomys bottae}. 

\hskip 0.5in 
To reconstruct trees, we used {\tt BEAST} with the GTR+I+G model (Figure 6).  
In this analysis, {\em  T. talpoides} and {\em T. bottae} were outgroups in the gopher data set, and 
{\em T. barbarae} and {\em T. minor} were outgroups in the louse data set.
With the full data sets,
application of {\tt ParaFit} to all pairwise distances gave $p = 0.001$.  In contrast, the dissimilarity method with the sampling fraction 0.0005 gave $p = 0.589$, not rejecting the null hypothesis (Table \ref{pvalues3}).  With the trimmed data sets, 
application of {\tt ParaFit} to all pairwise distances gave $p = 0.001$ and also the dissimilarity method with the sampling fraction = 0.0005 gave $p = 0.012$.
However, application of the dissimilarity method with MRCA pairs from MRCALink 
gave significant p-values, 
evidence to reject the null hypothesis for both 
the full data and trimmed gopher-louse data sets.   

\hskip 0.5in 
With the full data sets and the trimmed data sets, applying {\tt ParaFit} 
on all pairwise distances gave a significance at $p < 0.01$ for some 
sample fractions (Table \ref{pvalues3}). Applying {\tt ParaFit} to the MRCALink-derived matrix for 
these full and trimmed data sets also gave $p < 0.01$ for all sample 
fractions tried, indicating rejection of 
the null hypothesis. 
\vskip 0.2in

%\section*{Discussion}
\begin{center}
{\textsc{Discussion}}

{{\it Endophyte Codivergences with Hosts}}
\end{center}

\hskip 0.5in 
In this study we have taken a novel approach to investigate codivergences between hosts and symbionts 
(parasites), which differs from others \citep{Jackson,  Legendre, Page} in that it is a more direct comparison of historical 
cladogenesis events represented by inferred MRCA ages in ultrametric time trees in a way that avoids 
excessive weighting of deeper nodes compared with shallower nodes. The results indicate that, with 
relatively few exceptions, evolution of symbiotic epichlo\"e fungi largely tracked evolution of their grass 
hosts. These symbioses extend across the taxonomic range of the Pooideae, including the basal tribe 
Brachyelytreae, yet are restricted to this subfamily. Endophytes related to epichlo\"e (such as {\em 
Balansia} species) are known from other hosts, but the combination of very benign, often mutualistic, 
interactions and extremely efficient vertical transmission is known only in the Pooideae-epichloae 
system \citep{Clay}. The conclusion that the system is dominated by codivergence implies that this 
unusually intimate symbiotic system emerged coincidentally with the emergence of this important 
grass subfamily.

\hskip 0.5in 
In an earlier study comparing grass and endophyte evolutionary histories, the topological relationships 
of host tribes matched those of {\em Epichlo\"e} species with a mixed mode of transmission \citep{Schardl97}. 
However, no asexual lineages were included, and the possibility that some of the strictly sexual, 
horizontally transmitted species might also have a history of codivergence was not assessed. 
More importantly, the inference of codivergence was based on branching order, not relative timing of 
cladogenesis events. Although mirror phylogenies are suggestive of codivergence, it is the 
concomitance of corresponding cladogenesis events that defines codivergence \citep{Hafner}. 
Conversely, unless the codivergences correspond to actual speciation events (that is, isolation of 
populations into distinct gene pools), lineage sorting effects, species duplications, and incomplete taxon sampling can prevent $H$ and $P$ phylogenies 
from mirroring each other \citep{Page}. Therefore, we undertook the current study to assess 
codivergence by a more direct assessment of relative ages of corresponding cladogenesis events. 

\hskip 0.5in 
The null hypothesis was that relative ages of corresponding host and endophyte MRCAs were unrelated.
 If all sampled taxa were included, the null hypothesis was not rejected. This, however, was expected 
for the full data set for two reasons: (1) topology within the {\em E. typhina} clade bears no 
resemblance to 
that of the hosts, in keeping with the fact that {\em E. typhina} is a broad host-range species, 
and (2) some of the topological discordances in other clades strongly suggested occasional host jumps. 
Topological discordances tended to involve rarer associations. For example, {\em E. festucae} has 
only been identified in {\em K. cristata} once, but is very common in {\em Festuca} species. 
These features of the grass-endophyte system allowed for a rational basis for trimming trees of 
exceptions to assess support for codivergences in remaining taxa. When the {\em E. typhina}  clade was removed the significance of codivergence 
increased dramatically. Using all pairwise distances for dissimilarity analysis the null hypothesis was still not rejected, but when the method was applied to the MRCALink-derived data set, the null hypothesis was strongly rejected. Trimming other possible host jumps further decreased the p-values in all analyses, strongly supporting the conclusion that the relationships between host and endophyte trees had a degree of dependence.

\hskip 0.5in 
Various factors may cause a tendency for codivergence in evolution of these symbioses. 
For example, it may be much more likely for an endophyte to colonize a new host species that is 
closely related to its host of origin than a host that is more distantly related. A less likely 
(though not mutually exclusive) scenario would be speciation of hosts driven in part by adaptation to
 different symbionts. For example, considering that benefits of these endophytes are likely to be highly 
dependent on environmental conditions, geographical separation of the combined host-endophyte 
systems followed by different selective forces in the separate populations might eventually lead to a 
circumstance (after the populations merge again) in which sharing of endophytes is less beneficial to
 the hosts. Analogously to the problem of hybrid disadvantage, a ``hybrid symbiotum disadvantage'' 
may simultaneously promote evolution of genetic isolation for both the hosts and their endophytes 
\citep{Thompson}. 

\hskip 0.5in 
However, it is also possible that hosts will tend to benefit from endophytes that have adapted to 
related hosts, but will benefit far less or actually suffer detriment from endophytes adapted to 
distantly related hosts. So, for example, most {\em E. typhina, E. baconii} and {\em E. glyceriae} 
strains infecting their hosts cause complete suppression of seed production, thus eliminating a means 
of dispersal (seeds) as well as a means of genetic diversification (meiotic recombination) 
that could enhance survivability of host progeny in the face of changing environmental factors. 
In contrast, endophytes such as {\em E. festucae, E. elymi, E. brachyelytri}, and {\em E. amarillans} 
allow substantial seed production by producing their fruiting structures (stromata) on only a portion of 
the tillers of the infected plant \citep{Leuchtmann,  Schardl2003}. The situation with {\em E. bromicola} 
and {\em E. sylvatica} is more complicated because expression of host seeds or endophyte stromata 
depends on host and endophyte genotypes, but these species also have the potential to be far less 
damaging to their hosts than are {\em E. typhina, E. baconii} and {\em E. glyceriae}. It should be much more to 
the benefit of a host to maintain compatibility with the benign or mutualistic {\em Epichlo\"e} species than 
with the more antagonistic species. Indeed, {\em E. typhina} has a broad host range and clearly has not 
codiverged with those hosts \citep{Leuchtmann}, which is why this species and the asexual endophytes most closely 
related to it were removed in the trimmed data sets for analysis of codivergence. The question is
 whether the remaining narrower host-range endophytes have tended to codiverge with their hosts, 
and our results suggest that this is the case.

\hskip 0.5in 
The possibility that these symbiotic systems emerged with the origin of the grass subfamily is intriguing. 
\cite{Kellogg} postulated an early shift from shady to sunny habitats in subfamily Pooideae. According to 
this hypothesis, lineages derived following the split from the Brachyelytreae lineage moved into open habitats 
where, presumably, competition was less intense but solar radiation and drought were more prevalent. Additionally, 
such early colonizers may have been more conspicuous to potential herbivores. Protection from herbivory and 
drought are among the better documented effects of the epichloae \citep{Clay}. 
Kellogg's hypothesis adds perspective to our results suggesting that codivergence between cool-season 
grasses and their endophytes originated with the Pooideae. It is reasonable to suggest that these symbioses 
may have played an important role in this habitat shift and that the mutualistic tendencies that these fungi 
commonly impart to their hosts (drought tolerance, herbivore resistance, etc.) are a direct reflection of these 
new selective pressures faced in open habitats. Even the more antagonistic endophytes that severely restrict 
seed production would probably have exerted a strong effect on structuring emerging grassland communities. 
Following this habitat transition, these symbionts may have significantly enhanced host fitness, aiding the radiation 
of a highly successful and speciose grass subfamily. 

\vskip 0.4cm

\begin{center}
{{\it The MRCALink Method}}
\end{center}

\hskip 0.5in 
Our method involves direct comparison of MRCA ages rather than analysis of host and 
symbiont pairwise distance matrices, which is often used in studies of codivergence \citep{Legendre}. 
The pairwise distance approach is attractive in that it seems not to require inference of phylogenies, 
for which lineage sorting of preexisting polymorphisms, and imperfect genetic barriers between species, may give phylogenies that inaccurately represent 
histories of speciation \citep{Page}. Nevertheless, there is a true phylogeny of species, and each pair 
of leaves (extant taxa) for which a pairwise distance is calculated represents the node in that phylogeny
 that is their MRCA. Codivergence implies that the MRCA of a host leaf pair occurred at the same time 
as the MRCA of a symbiont leaf pair. If we test all host leaf pairs and corresponding symbiont leaf pairs
 in order to derive a statistical test of the null hypothesis that there is no significant relationship between 
host and symbiont MRCA times, a problem emerges in the number of times each MRCA is sampled. 
In the case of symmetrical and mirror (balanced) $H$ and $P$ trees, the frequency of 
node sampling ($= 2^{2m-2}$ where $m$ is the level in the tree) increases exponentially as one goes deeper into the tree. 
For example, suppose we have the situation that both trees are balanced and the tree topologies are 
congruent. Then the MRCA pair constituting the roots of both trees is sampled $(n/2)^2$ times 
(where $n$ is the number of taxa in each tree), whereas the MRCA pairs from corresponding tip 
clades are sampled only once. In this case, the number of times a MRCA pair is sampled is $(n_1 - 1)^2+ 
(n_2 - 1)^2$  where $n_1$ and $n_2$ are the numbers of descendants from the MRCAs in $H$ 
and $P$ trees, respectively. To examine the more generalized case of $H$ and $P$ trees that may not 
be balanced or congruent, 10,000 random $H$ and $P$ trees were generated with 25 taxa in 
PAML by the BDP with birth rates = 0.5, death rates = 0.5 and mutation rates = 100. 
Then, for each pair of $H$ and $P$ trees we identified all corresponding leaf pairs and counted how 
many times each MRCA pair was identified. The maximum number of times any MRCA pair was sampled averaged 111 in the 10,000 simulations.

\hskip 0.5in 
In this study we introduce the MRCALink algorithm to specifically identify valid $H$ and $P$ MRCA pairs 
to compare divergence times, and to avoid repeated sampling of any MRCA pairs. 
The aforementioned analysis suggests that this method of sampling nodes is a substantial improvement 
over the use of complete pairwise distance matrices. However, the MRCA method is affected by incongruence of $H$ and $P$ trees. As shown in Figure 1, the nodes (MRCAs) of subtrees 
where such incongruences exist will tend to be sampled more than the nodes in congruent subtrees, 
even though each node pair is sampled no more than once. This may be why, compared to the use of all pairwise distances, application of MRCALink is more sensitive to dependence between the trees being compared by the dissimilarity method in which the ML tree pair is compared to pairs of randomly generated trees of equal length. This may be considered a benefit of the MRCALink method, but further research into this behavior and possible modifications of the method are required to fully assess how this characteristic of the method affects inference about tree dependence. 

\hskip 0.5in 
Our method described in this paper does not find the 
minimum set of non-codiverging taxon pairs.  It would be interesting to
find an efficient algorithm to find the minimum set of non-codiverging 
taxon pairs from $H$ and $P$.  
\vskip 0.4cm

\begin{center}
{{\it Dissimilarity Method}}
\end{center}

\hskip 0.5in 
The p-values obtained via the dissimilarity 
method for all pairwise distances tend be larger than the p-values for MRCA pairs. 
This is because (1) analysis of all pairwise distances will generate a bias 
in favor of non-codivergence, 
(2) the $S$ distribution is the sum of absolute values of 
differences between distances for each pair of taxa, and (3) from the Computation Results in the Analyses of Codivergence subsection above, if we take all pairwise distances, then we observed that 
MRCA pairs tend to be more frequently sampled in highly correlated trees than in poorly correlated trees. 
Since MRCA pairs for highly correlated $T_H$ and $T_P$
seem to be more frequently sampled among all pairwise distances than  MRCA pairs for less correlated trees, the 
$S$ value for $T_H$ and $T_P$ for all pairwise distances is overestimated 
and thus an estimated p-value obtained via the dissimilarity 
method with all pairwise distances is likely to be higher than the actual
 p-value
(see Tables \ref{pvalues1} -- \ref{pvalues3}). 

\hskip 0.5in 
Generating random trees by the BDP tends to produce 
trees with long interior branches. 
\cite{Yang2} suggest that taking incomplete species sampling 
into account generates more realistic trees.
Thus we used {\tt evolver} to generate random ultrametric trees with specified
sampling fractions.  \cite{Aris} suggest using a 
sampling fractions in $(0, 0.001)$. However, 
\cite{Aris} also note that the sampling fraction is known to 
affect the topology of random trees, thus 
may affect divergence time of nodes.   
Therefore, in this study we took four different sampling fractions 0.0005, 0.001, 0.5 and 1, though we consider the smallest sampling fraction to be the most biologically relevant \citep{Aris}.

\hskip 0.5in 
We have used random trees to estimate the measure between two trees. 
Currently we use the BDP with a specified sampling fraction. When we measure a dissimilarity between the host and parasite trees in the space of trees, we used a heuristic method because measuring the exact distance between two trees in the space of trees requires  exponential time in terms of the number of taxa \citep{treespace}.  However, \cite{amenta} recently developed a method to approximate a distance between trees in the space of trees efficiently, which could be a reasonable alternative to our method.   
\vskip 0.4cm

\begin{center}
{{\it ParaFit Method}}
\end{center}

\hskip 0.5in 
From results in 
Table \ref{pvalues4} it seems that some of the p-values obtained via {\tt ParaFit} with MRCA results are higher 
than the p-values via {\tt ParaFit} with all pairwise distances.  
This may be due to the use of PCA prior to {\tt ParaFit}. 
In the process of removing some of the entries in each distance matrix,  
we cut off more small signals in the data. 

However, we note that p-values for sub-optimal trees via {\tt ParaFit} 
differ from  p-values for the ML tree and in some instances
have much larger p-value than p-values from the ML 
trees (e.g., Table \ref{pvalues4}). 
Also there are large differences between p-values from the ML trees and 
sub-optimal trees for the plant-endophyte data sets. 
For the ML trees  {\tt ParaFit} returns p-values of 0.001 for all data
sets.  However, with some of the sub-optimal trees {\tt ParaFit} returns
higher p-values not rejecting the null hypotheses. These sub-optimal trees are sampled from the distribution with the given data and model via MCMC. 
 
\hskip 0.5in 
Comparing the dissimilarity method and ParaFit method, both estimate 
the p-values by sampling random trees or random matrices, respectively.  {\tt ParaFit}
takes two distance matrices, then estimates the p-value by the 
permutation test.  The dissimilarity 
method estimates the p-value by sampling random ultrametric trees from
the BDP with a given sampling fraction known to
be biologically meaningful.
Also {\tt ParaFit} does not include constraints that these
random matrices are distance matrices and coming from trees.
On the other hand, the dissimilarity
method includes constraints that these random samples are ultrametric trees. Therefore,
these biologically constraints added in the dissimilarity method should result in more
biologically meaningful p-values. Also, it is interesting that the p-values computed from
sub-optimal trees with $T_4$ data set did not give strong evidence for rejecting the hypothesis, but the ML method did. This suggests that the ParaFit method may be very sensitive to parametric assumptions and/or be unstable.

\vskip 0.2in

%\vskip 0.4cm
\begin{center}
%\noindent{Acknowledgments}
\noindent{\textsc{Acknowledgments}}
\end{center}

\hskip 0.5in 
We thank  Adrian Leuchtmann (ETH Z\"urich, Switzerland) for providing biological materials, Walter Hollin (University of Kentucky) for technical support,  Stephane Aris--Brosou for very useful conversation, and  Mark Hafner for providing the gopher-louse data sets. Also we would like to thank Roderic Page,  Francois-Joseph Lapointe, Michael Charleston, Jack Sullivan, and an anonymous referee for very useful comments to improve this paper. This work was supported by National Science Foundation grant DEB--9707427, National Institutes of Health grant KY--INBRE P20 RR16481, U.S.Department of Agriculture grant CSREES Grant 2005-34457-15712, and the Harry E. Wheeler Endowment to the University of Kentucky Department of Plant Pathology. This is publication 08--12--034 of the Kentucky Agricultural Experiment Station, published with approval of the director. % Sequence alignments and trees reconstructed via {\tt MrBayes} and {\tt BEAST} of plants and endophytes have been submitted to TreeBASE.

\pagebreak
\renewcommand{\refname}{\begin{center}{\sc References}\end{center}}

\pagebreak
\appendix
  \begin{center}
    {\bf APPENDIX}
  \end{center}

%\section{The MRCALink Algorithm}\label{appendix}

\begin{center}
{\textsc{The MRCALink Algorithm}}
\end{center}

\hskip 0.5in 
Given a set of host taxa $H$ and a set of symbiont taxa $P$ (``parasites,'' in keeping with other literature in the field), 
there is a map called $L: H \to P$ such that a host $A \in H$ has a parasite or symbiont $L(A) \in P$. 
Define $MRCA(A, B)$ to be the node representing the Most Recent Common Ancestor (MRCA) of leaves $A$ 
and $B$.  
\begin{algorithm}[The MRCALink Algorithm]\label{sample}

\noindent
\begin{itemize}
\item {\bf Input} a set of host taxa $H$, a set of parasite taxa $P$, a $H$ tree $T_H$, and a  $P$ tree $T_P$ where $n_1$ is the number of taxa in $H$ and $n_2$ is the number of taxa in $P$. 
\item {\bf Output} a set of MRCA pairs of host taxa and parasite taxa.
\item {\bf Algorithm}
\begin{tabbing}
\quad \= \quad \= \quad \= \quad \= \quad \kill
Assign each node a unique number from $1$ to $2n_1-1$ in the 
host tree and\\ a unique number from $1$ to $2n_2-1$ in the 
parasite tree
such that a node $i$ is \\ancestral to a node $j$.\\ 
Let $U$ be a set of pairs of a 
pair of taxa in $H$ and a pair of taxa in $P$, initially empty.\\
{\bf for }$(i$ from $n_1+1$ to $2n_1-1)$ {\bf do}\{\\
\> Set $X_i = l_i \times r_i$ where $l_i$ is the set of all left-descendants 
of $i$,\\ 
\> \> and where $r_i$ is the set of all right-descendants of $i$.\\
\> /* This is just another way of saying $X_i$ is all such pairs of one leaf 
\\ \> \>from the left and one from the right. */\\
\> {\bf while} $(X_i \not = \emptyset)$ {\bf do}\{\\
\> \> Choose $x = MRCA(a, b) \in X_i$ and identify $y_j = MRCA(L(a),L(b))$ for each\\
\> \> \> distinct $L(a)$ and $L(b)$.\\
\> \> Remove $x$ from $X_i$.\\
\> \> {\bf for} $($each distinct $y_j)$ {\bf do}\{\\
\> \> \> {\bf if} $(MRCA(x, y_j) \not \in U)$ {\bf do}\{\\
\> \> \> \> $U \leftarrow  U \cup MRCA(x, y_j)$.\\
\> \> \> \}\\
\> \>\}\\
\> \}\\
\}\\
Output $U$.\\
\end{tabbing}
\end{itemize}
\end{algorithm}

\pagebreak

\pagebreak

\begin{table}[ht!]
\begin{center}
\caption{Hosts and symbionts. All listed taxa, as well as trimmed taxon sets $T_1$--$T_4$, were assessed for probability of codivergence.}\label{table1}
%\caption{Hosts and symbionts.}\label{table1}
\begin{tabular}{llllll}%\\

\hline

&

&

\multicolumn{4}{l} {\small Included in:} \\
\hline

{\small Grasses }&

{\small Endophytes}&

{\small $T_1$}&

{\small $T_2$}&

{\small $T_3$}&

{\small $T_4$}\\
\hline

\textit{{\small Brachyelytrum erectum }}{\small  (root)}&

\textit{{\small Epichlo\"e brachyelytri }}{\small  (root)}&

+&

+&

+&

+\\

\textit{{\small Brachypodium sylvaticum }}&

\textit{{\small Epichlo\"e sylvatica }}{\small  200751}&

{\small +}&

--&

+&

--\\

\textit{{\small Echinopogon ovatus}}&

\textit{{\small Neotyphodium aotearoae }}{\small  829}&

{\small +}&

--&

+&

--\\

\textit{{\small Calamagrositis villosa}}&

\textit{{\small Epichlo\"e baconii }}{\small  200745}&

{\small +}&

+&

+&

+\\

\textit{{\small Agrostis tenuis}}&

\textit{{\small Epichlo\"e baconii }}{\small  200746}&

{\small +}&

+&

+&

+\\

\textit{{\small Agrostis hiemalis}}&

\textit{{\small Epichlo\"e amarillans }}{\small  200744}&

{\small +}&

+&

+&

+\\

\textit{{\small Sphenopholis obtusata}}&

\textit{{\small Epichlo\"e amarillans }}{\small  200743}&

{\small +}&

+&

+&

+\\

\textit{{\small Koeleria cristata}}&

\textit{{\small Epichlo\"e festucae }}{\small  1157}&

{\small +}&

+&

--&

--\\

\textit{{\small Lolium }}{\small sp. \ P4074}&

\textit{{\small Neotyphodium }}{\small  sp.\ FaTG2 \ 4074}&

{\small +}&

+&

+&

+\\

\textit{{\small Lolium }}{\small sp. \ P4078}&

\textit{{\small Neotyphodium }}{\small  sp.\ FaTG3 \ 4078}&

{\small +}&

+&

+&

+\\

\textit{{\small Lolium arundinaceum}}&

\textit{{\small Neotyphodium coenophialum }}{\small  19}&

{\small +}&

+&

+&

+\\

\textit{{\small Lolium multiflorum}}&

\textit{{\small Neotyphodium occultans }}{\small  999}&

{\small +}&

+&

+&

+\\

\textit{{\small Lolium edwardii}}&

\textit{{\small Neotyphodium typhinum }}{\small  989}&

--&

--&

--&

--\\

\textit{{\small Lolium perenne}}&

\textit{{\small Epichlo\"e typhina }}{\small  200736}&

--&

--&

--&

--\\

\textit{{\small Lolium perenne}}&

\textit{{\small Neotyphodium lolii }}{\small  135}&

{\small +}&

+&

--&

--\\

\textit{{\small Festuca rubra}}&

\textit{{\small Epichlo\"e festucae }}{\small  90661}&

{\small +}&

+&

+&

+\\

\textit{{\small Festuca longifolia}}&

\textit{{\small Epichlo\"e festucae }}{\small  28}&

{\small +}&

+&

+&

+\\

\textit{{\small Holcus mollis}}&

\textit{{\small Epichlo\"e }}{\small  sp. 9924}&

{\small +}&

+&

+&

+\\

\textit{{\small Hordelymus europaeus}}&

\textit{{\small Neotyphodium }}{\small  sp. 362}&

{\small +}&

+&

+&

+\\

\textit{{\small Bromus ramosus}}&

\textit{{\small Epichlo\"e bromicola }}{\small  201558}&

{\small +}&

+&

+&

+\\

\textit{{\small Bromus erectus}}&

\textit{{\small Epichlo\"e bromicola }}{\small  200749}&

{\small +}&

+&

+&

+\\

\textit{{\small Bromus purgans}}&

\textit{{\small Epichlo\"e elymi }}{\small  1081}&

{\small +}&

+&

--&

--\\

\textit{{\small Hordeum brevisubulatum }}&

\textit{{\small Neotyphodium }}{\small  sp. 3635}&

{\small +}&

+&

+&

+\\

\textit{{\small Elymus canadensis }}&

\textit{{\small Epichlo\"e elymi }}{\small 201551}&

{\small +}&

+&

+&

+\\

\textit{{\small Glyceria striata}}&

\textit{{\small Epichlo\"e glyceriae }}{\small  200755}&

{\small +}&

+&

+&

+\\

\textit{{\small Achnatherum inebrians}}&

\textit{{\small Neotyphodium gansuense }}{\small  818}&

{\small +}&

+&

+&

+\\
\hline
\end{tabular}
\end{center}
\end{table}

%\pagebreak

\small{
\ctable[
cap = Primer list,
caption = {Primer list. Oligonucleotide primers for amplification and sequencing of plant cpDNA intron and intergenic regions.}, %: Oligonucleotide primers for amplification and sequencing of plant cpDNA intron and intergenic regions.
label = table2,
]{llll}{
\tnote[a]{Primers used in PCR and sequencing.}
\tnote[b]{Internal primers used in sequencing only.}
}{
\FL
{Primer}&
\uline{{Region}}&
{\textsf{Sequence (5'--3')}}&
\uline{{Orientation}}\LL
{B48557}$^{\mathrm{{a}}}$  & %\footnote{We denote $a$ Primers used in PCR and sequencing and $b$ Internal primers used in sequencing only.}&
{trnT-trnLspacer}&
{CATTACAAATGCGATGCTCT}&
{downstream}\NN
{A49291}$^{\mathrm{{a}}}$&
{trnT-trnLspacer}&
{TCTACCGATTTCGCCATATC}&
{upstream}\NN
{B49317}$^{\mathrm{{a}}}$&
{trnL intron}&
{CGAAATCGGTAGACGCTACG}&
{downstream}\NN
{A49855}$^{\mathrm{{a}}}$&
{trnL intron}&
{GGGGATAGAGGGACTTGAAC}&
{upstream}\NN
{B49873}$^{\mathrm{{a}}}$&
{trnL-trnF spacer}&
{GGTTCAAGTCCCTCTATCCC}&
{downstream}\NN
{A50272}$^{\mathrm{{a}}}$&
{trnL-trnF spacer}&
{ATTTGAACTGGTGACACGAG}&
{upstream}\NN
{trnTtrnL766-747u}$^{\mathrm{{b}}}$&
{trnT-trnL spacer}&
{GAATCATTGAATTCATCACT}&
{upstream}\NN
{trnTtrnL359-340u}$^{\mathrm{{b}}}$&
{trnT-trnL spacer}&
{TATTAGATTATTCGTCCGAG}&
{upstream}\NN
{trnTtrnL306-325d}$^{\mathrm{{b}}}$&
{trnT-trnLspacer}&
{GGAATTGGATTTCAGATATT}&
{downstream}\NN
{trnTtrnL601-621d}$^{\mathrm{{b}}}$&
{trnT-trnL spacer}&
{AATATCAAGCGTTATAGTAT}&
{downstream}\NN
{P37.trnTtrnL.359-340u}$^{\mathrm{{b}}}$&
{trnT-trnL spacer}&
{TATTAGATTTCTCCTCTGAG}&
{upstream}\NN
{P56.trnTtrnL.378-398d}$^{\mathrm{{b}}}$&
{trnT-trnL spacer}&
{TAAGACGGGAGGTGGG}&
{downstream}\NN
{P56.trnTtrnL.398-378u}$^{\mathrm{{b}}}$&
{trnT-trnL spacer}&
{CTCCCCCACCTCCCGTCTTA}&
{upstream}\NN
{P57trnTtrnL556-576d}$^{\mathrm{{b}}}$&
{trnT-trnL spacer}&
{GTCATAGCAAATAAAATTGC}&
{downstream}\NN
{P2772.trnTtrnL.306-325d}$^{\mathrm{{b}}}$&
{trnT-trnL spacer}&
{CTAATTGGATTTTAGATATT}&
{downstream}\NN
{Bromus.trnTtrnL.177-197d}$^{\mathrm{{b}}}$&
{trnT-trnL spacer}&
{TTGATATGCTTAACTATAGG}&
{downstream}\NN
{Bromus.trnTtrnL.197-177u}$^{\mathrm{{b}}}$&
{trnT-trnL spacer}&
{CCTATAGTTAAGCATATCAA}&
{upstream}\NN
{Bromus.trnTtrnL.357-376d}$^{\mathrm{{b}}}$&
{trnT-trnL spacer}&
{GCGTTATAGTATAATTTTG}&
{downstream}\NN
{Bromus.trnTtrnL.376-357u}$^{\mathrm{{b}}}$&
{trnT-trnL spacer}&
{CAAAATTATACTATAACGC}&
{upstream}\NN
{trnLintron285-303d}$^{\mathrm{{b}}}$&
{trnL intron}&
{CATAGCAAACGATTAATCA}&
{downstream}\NN
{trnLintron303-285u}$^{\mathrm{{b}}}$&
{trnL intron}&
{TGATTAATCGTTTGCTATG}&
{upstream}\NN
{trnLtrnF.77-97d}$^{\mathrm{{b}}}$&
{trnL-trnF spacer}&
{TTTAAGATTCATTAGCTTTC}&
{downstream} \LL
}
}

\pagebreak

\begin{table}[ht!] 
\begin{center}
\caption{The p-values obtained by applying the dissimilarity method to all pairwise distances (noted by ALL) and to the MRCALink-derived matrix (noted by MRCA) for full and $T_1$ -- $T_4$ plant and endophyte data sets (see Table \ref{table1} for the data sets).  SF means a sampling fraction.}%The p-values obtained by applying the dissimilarity method to all pairwise distances (noted by ALL) and to the MRCALink-derived matrix (noted by MRCA) for full and $T_1$ -- $T_4$ plant and endophyte data sets (see Table \ref{table1} for the data sets).  SF means a sampling fraction.} 
\label{pvalues1}
%\caption{Averages of computational experiences} 
\begin{tabular}{llccccl} \hline
Method & Data & SF = 0.0005 & SF = 0.001 &     SF =   0.5&      SF =   1.0 \\\hline
ALL &Full&   0.784 & 0.783 & 0.677 & 0.374  \\
MRCA &Full&  0.123 & 0.123 & 0.081 & 0.039 \\
ALL&$T_1$ &  0.117 & 0.115 & 0.035 & 0.009 \\
MRCA &$T_1$ & $<0.001$ & $<0.001$  & $<0.001$  &  $<0.001$  \\
ALL&$T_2$ &  0.093 & 0.085 & 0.027 & 0.012 \\
MRCA &$T_2$ & $<0.001$  &  $<0.001$  &  $<0.001$  &  $<0.001$ \\
ALL&$T_3$ & 0.064  & 0.061 & 0.017 & 0.005 \\
MRCA &$T_3$ & $<0.001$  &  $<0.001$   & $<0.001$  &  $<0.001$  \\
ALL&$T_4$ & 0.018  & 0.020  & 0.005 & 0.002    \\
MRCA &$T_4$ & $<0.001$  &  $<0.001$  &  $<0.001$ &   $<0.001$  \\\hline
\end{tabular}
\end{center}
\end{table}

\pagebreak

\begin{table}[ht!] 
\begin{center}
\caption{The p-values obtained using the dissimilarity method with sub-optimal trees with 26 full and $T_1$ -- $T_4$ plant and endophyte data sets (all taxa listed in Table \ref{table1}) via the Bayesian MCMC method in {\tt BEAST}.  ALL means the dissimilarity method with all pairwise distances, and MRCA means the dissimilarity method with the MRCALink-derived matrix.  SF means a sampling fraction.  Each sampled tree is assigned a number from 1 to 3 to distinguish it from the others.}%The p-values obtained using the dissimilarity method with sub-optimal trees with 26 full and $T_1$ -- $T_4$ plant and endophyte data sets (all taxa listed in Table \ref{table1}) via the Bayesian MCMC method.  ALL means the dissimilarity method with all pairwise distances and MRCA means the dissimilarity method with the MRCALink-derived matrix.  SF means a sampling fraction.  Each sampled tree is assigned number from 1 to 3 to distinguish it from the others.} 
\label{pvalues2}
%\caption{Averages of computational experiences} 
\begin{tabular}{lllccccl} \hline
Method& Data & sample number& SF = 0.0005 & SF = 0.001 &    SF =    0.5&   SF =      1.0 \\\hline
ALL & Full &sample 1 & 0.700 & 0.686 & 0.466 & 0.294  \\
MRCA &Full &sample 1 & 0.011 & 0.011 & 0.003 & 0.002  \\
ALL &Full &sample 2  & 0.474 & 0.483 & 0.245 & 0.119  \\
MRCA &Full &sample 2 & 0.064 & 0.064 & 0.025 & 0.014  \\
ALL &Full &sample 3  & 0.684 & 0.683 & 0.450 & 0.262 \\
MRCA &Full &sample 3 & 0.193 & 0.190 & 0.102 & 0.061 \\
ALL & $T_1$ &sample 1& 0.451 & 0.448 & 0.236 & 0.115 \\
MRCA &$T_1$ &sample 1& $<0.001$ & $<0.001$ & $<0.001$ & $<0.001$ \\
ALL & $T_1$&sample 2 & 0.029 &  0.033 & 0.005 & $< 0.001$  \\
MRCA & $T_1$&sample 2& $<0.001$ & $<0.001$ & $<0.001$ & $<0.001$ \\
ALL & $T_1$&sample 3 & 0.006 &  0.007 & $<0.001$ & $<0.001$\\
MRCA &$T_1$ &sample 3& $<0.001$ & $<0.001$ & $<0.001$ & $<0.001$ \\
ALL & $T_2$ &sample 1& 0.346 &  0.355 & 0.190 &  0.097 \\
MRCA &$T_2$ &sample 1& $<0.001$ & $<0.001$ & $<0.001$ & $<0.001$ \\
ALL & $T_2$&sample 2 & 0.355 &  0.360 &  0.184 & 0.099 \\
MRCA & $T_2$&sample 2& $<0.001$  & $<0.001$ & $<0.001$ &  $<0.001$ \\
ALL & $T_2$&sample 3 & 0.084  & 0.079  & 0.022 &  0.010 \\
MRCA &$T_2$ &sample 3& $<0.001$ & $<0.001$ & $<0.001$ & $<0.001$ \\
ALL & $T_3$ &sample 1& 0.070 & 0.067 & 0.020 & 0.007 \\
MRCA &$T_3$ &sample 1& $<0.001$ & $<0.001$ & $<0.001$ & $<0.001$ \\
ALL & $T_3$&sample 2 & 0.030 &  0.029  & 0.007  & 0.030 \\
MRCA & $T_3$&sample 2& $<0.001$ & $<0.001$ & $<0.001$ & $<0.001$  \\
ALL & $T_3$&sample 3 & 0.132&  0.138 & 0.050 & 0.021 \\
MRCA &$T_3$ &sample 3& $<0.001$ & $<0.001$ & $<0.001$ & $<0.001$  \\
ALL & $T_4$ &sample 1& 0.106 & 0.103 &  0.039 &  0.015 \\
MRCA &$T_4$ &sample 1& $<0.001$ & $<0.001$ & $<0.001$ & $<0.001$  \\
ALL & $T_4$&sample 2 & 0.024 & 0.026 &  0.007  & 0.002 \\
MRCA & $T_4$&sample 2& $<0.001$ & $<0.001$ & $<0.001$ & $<0.001$  \\
ALL & $T_4$&sample 3 & 0.017 & 0.016 & 0.006 &   0.002 \\
MRCA &$T_4$ &sample 3& $<0.001$ & $<0.001$ & $<0.001$ & $<0.001$  \\
\hline
\end{tabular}
\end{center}
\end{table}

\newpage

\begin{table}[ht!]
\begin{center}
\caption{The p-values obtained through {\tt ParaFit} to all pairwise distances (ALL) and to the MRCALink-derived matrix (MRCA) for the sub-optimal trees of 26 full and $T_1$ -- $T_4$ plant and endophyte data sets.  The sub-optimal trees were chosen as samples of the most common likelihood trees.  This was done by choosing samples with tree likelihoods close to the mean of all 10,000 tree likelihoods. For each data set, the three sampled sub-optimal trees were arbitrarily assigned a number from 1 to 3.}%The p-values obtained through {\tt ParaFit} to all pairwise distances (ALL) and to the MRCALink-derived matrix (MRCA) for the sub-optimal trees of 26 full and T1 Ð T4 plant and endophyte data sets.  The sub-optimal trees were chosen as samples of the most common likelihood trees.  This was done by choosing samples with tree likelihoods close to the mean of all 10,000 tree likelihoods. For each data set, the three sampled sub-optimal trees were arbitrarily assigned a number from 1 to 3.}
\label{pvalues4}
\begin{tabular}{llcccll} \hline
Method &Data & Sample 1 & Sample 2 & Sample 3 \\\hline
ALL &Full & 0.146 & 0.007 & $<0.001$  \\
MRCA & Full &0.097 & 0.014 & 0.020 \\
ALL &$T_1$ &0.011 & 0.003  &0.006 \\
MRCA &$T_1$&0.023 & 0.029 & 0.115 \\
ALL &$T_2$ & $<0.001$ & $<0.001$ & 0.004\\
MRCA &$T_2$& $<0.001$ & 0.033 & 0.033 \\
ALL &$T_3$ &0.003 & $<0.001$ & 0.002 \\
MRCA &$T_3$&0.015 & $<0.001$ & 0.017\\
ALL &$T_4$ &0.046 & 0.046 & 0.042\\
MRCA &$T_4$&0.108 & 0.084 & 0.066 \\\hline
\end{tabular}
\end{center}
\end{table}

\pagebreak

\begin{table}[ht!] 
\begin{center}
\caption{The p-values obtained using the dissimilarity method for the gopher and louse data. The full data set includes all hosts and parasites from \citep{Hafner}, whereas for the trimmed data set data were removed for the gophers and lice which are from \citep{huelsenbeck}. ALL means the dissimilarity method with all pairwise distances and MRCA means the dissimilarity method with the MRCALink-derived matrix.}%The p-values obtained using the dissimilarity method for the gopher and louse data. The full data set includes all hosts and parasites from \citep{Hafner}, whereas for the trimmed data set data were removed for the gophers and lice which are from \citep{huelsenbeck}. ALL means the dissimilarity method with all pairwise distances and MRCA means the dissimilarity method with the MRCALink-derived matrix.} 
\label{pvalues3}
%\caption{Averages of computational experiences} 
\begin{tabular}{llccccl} \hline
Method& Data & SF = 0.0005 & SF = 0.001 &    SF =    0.5&     SF =    1.0 \\\hline
ALL &Full &    0.589 & 0.577  & 0.394 & 0.248 \\
MRCA &Full &    0.002 & 0.003  & $<0.001$ & $<0.001$ \\
ALL &Trimmed & 0.012 & 0.015   & 0.006  & 0.003  \\
MRCA &Trimmed & $<0.001$ & $<0.001$  & $<0.001$   & $<0.001$  \\\hline
\end{tabular}
\end{center}
\end{table}
\pagebreak

\begin{center}
{\it Legends to Figures}
\end{center}

%\begin{figure}[ht!]
%\begin{center}
   %  \includegraphics[scale= 0.6]{Figure1MRCALink.eps}
% \end{center}
%\caption{\doublespacing 
\vskip 0.2in

\noindent
{\bf Figure 1.}  
Simple examples of congruent and incongruent $H$ and $P$ trees, where $H$ is a set of plant or animal hosts and $P$ is a set of their symbionts or parasites, demonstrating the relationships of MRCA (most recent common ancestor) pairs to their corresponding $H$ and $P$ taxon pairs. In an ultrametric time tree, the distance between any two taxa is twice the age of their 
MRCA. In each tip clade a MRCA uniquely relates two taxa, but a MRCA deeper in the tree relates multiple taxon pairs. Therefore, pairwise distance matrices represent tip clade MRCAs once, but deeper MRCAs multiple times. The effect on sampling MRCA pairs is illustrated below each tree. For both congruent and incongruent pairs of $H$ and $P$ trees, comparison of pairwise distance matrices gives greater representation to pairs that include deeper MRCAs than to pairs of shallower MRCAs. The MRCALink algorithm samples corresponding $H$ and $P$ MRCA pairs only once. %}\label{fig2}
%\end{figure}

%\begin{figure}[ht!]
%\begin{center}
    % \includegraphics[scale= 0.6]{Figure2HostML.eps}
% \end{center}
%\caption{\doublespacing 
\vskip 0.2in

\noindent
{\bf Figure 2.}
Majority rule consensus tree with average branch lengths from Bayesian with General Time-Reversible plus gamma distribution (GTR+G) analysis of cpDNA intron and intergenic sequences from pooid grasses. The first number in each pair is a posterior probability and the second number in each pair indicates bootstrap support percentages (if $> 50$\%) obtained by 1000 maximum parsimony searches with branch swapping.  Currently accepted tribes are indicated at right. Full taxon names are given in Table \ref{table1}.%}\label{fig4}
%\end{figure}

%\begin{figure}[ht!]
%\begin{center}
     %\includegraphics[scale= 0.6]{Figure3HostML.eps}
% \end{center}
%\caption{\doublespacing  

\vskip 0.2in

\noindent
{\bf Figure 3.}
Majority rule consensus tree with average branch lengths from Bayesian (GTR+G) analysis from mainly intron sequences of endophyte $tefA$ and $tubB$ genes. Branches are labeled with posterior probability followed by bootstrap support percentage (if over 50\%) obtained by 1000 maximum parsimony searches with branch swapping. Host species are indicated in parentheses. Full taxon names are given in Table \ref{table1}. The LAE ({\em Lolium}-associated endophyte) clade is labeled.%}\label{figendo}
%\end{figure}

%\begin{figure}[ht!]
%\begin{center}
     %\includegraphics[scale= 0.6]{Figure4PlantEndo.eps}
% \end{center}
%\caption{\doublespacing  
\vskip 0.2in

\noindent
{\bf Figure 4.}
Ultrametric maximum likelihood (ML) time trees for host grasses and their endophytes. Hosts and their endophytes are indicated by dashed lines. Full taxon names are given in Table \ref{table1}. Numeric values on nodes represent their posterior probabilities estimated by {\tt BEAST}. The individual node posterior probabilities were calculated from nodes with a posterior probability greater than $0.5$, by {\tt Tree Annotator} in the {\tt BEAST} package. Labels preceding endophyte names indicate $H$ and $P$ pairs retained in trimmed data sets $T_1$--$T_4$. The LAE ({\em Lolium}-associated endophyte) clade is labeled.%}\label{fig1}
%\end{figure}

% \begin{wrapfigure}{r}{100mm}
%\begin{figure}[ht!]
%\begin{center}
     %\includegraphics[scale= 0.6]{BeastPlots2.eps}
%\end{center} 
%\caption{\doublespacing 
\vskip 0.2in

\noindent
{\bf Figure 5.}
Plots of relative MRCA ages of hosts (H) and their corresponding endophytes (P) identified by the MRCALink algorithm from ultrametric ML trees for the full dataset or trimmed datasets $T_1$ -- $T_4$, as indicated (see Table 1). %}\label{fig3}
%\end{figure}
%\end{wrapfigure}
%\pagebreak

%\pagebreak

%\begin{figure}[ht!]
%\begin{center}
     %\includegraphics[scale= 0.8]{Figure8GLPETE.eps}
% \end{center}
%\caption{\doublespacing 
\vskip 0.2in

\noindent
{\bf Figure 6.}
Ultrametric ML time trees for gopher and louse data sets \citep{Hafner} constructed via {\tt BEAST}. Hosts and their parasites are indicated by connecting dashed lines. Genera: {\em O. = Orthogeomys, Z. = Zygogeomys, P. = Pappogeomys, C. = Cratogeomys, G. = Geomys, T. = Thomomys, Gd. = Geomydoecus, Td = Thomomydoecus.}%}
%\label{GLfig}
%\end{figure}

\pagebreak

{\bf Figure 1.}%\ref{fig2}}.

\begin{figure}[ht!]
\includegraphics[scale= 0.9]{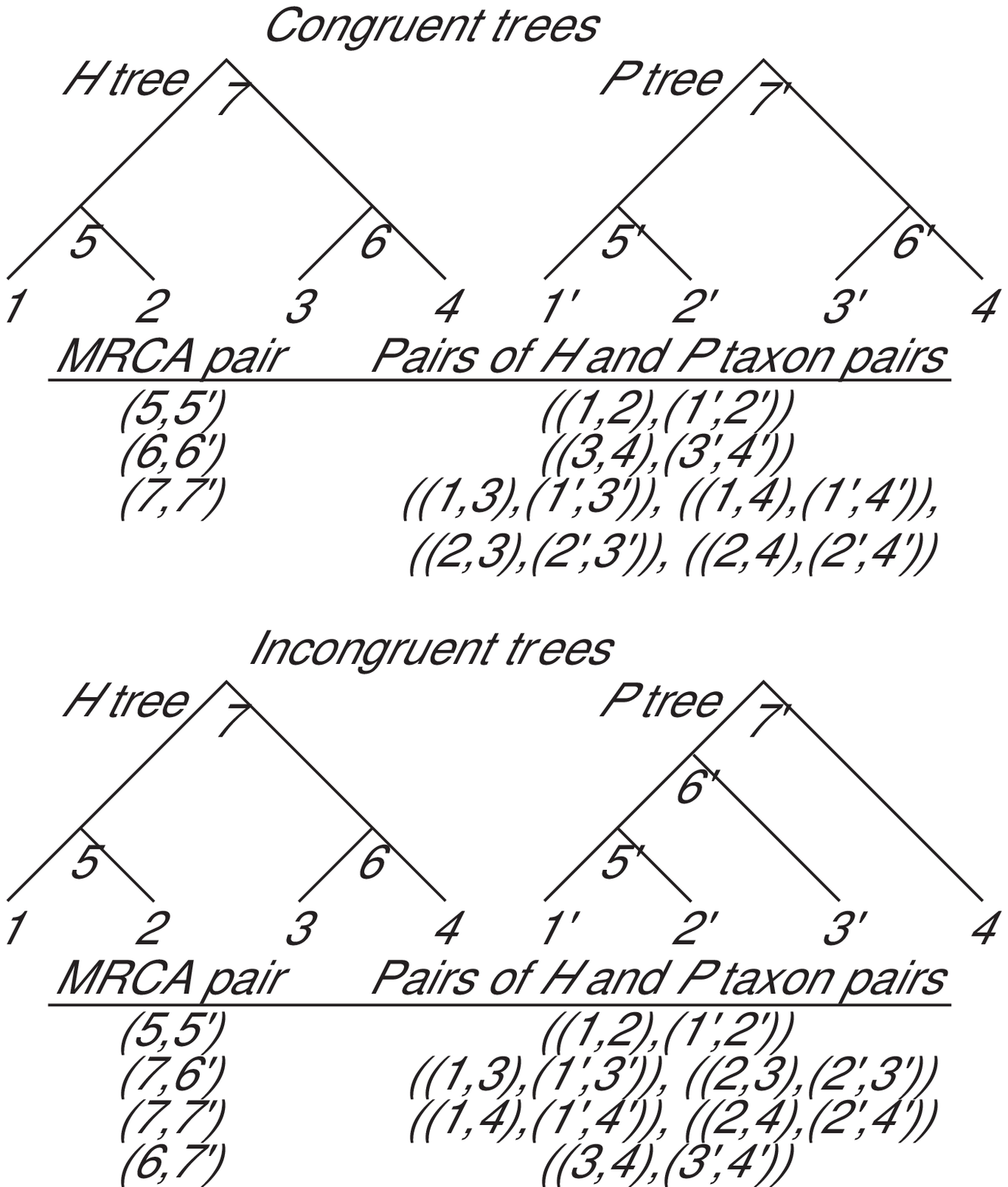}
\end{figure}

\pagebreak

{\bf Figure 2.}%\ref{fig4}}.

\begin{figure}[ht!]
\includegraphics[scale= 0.8]{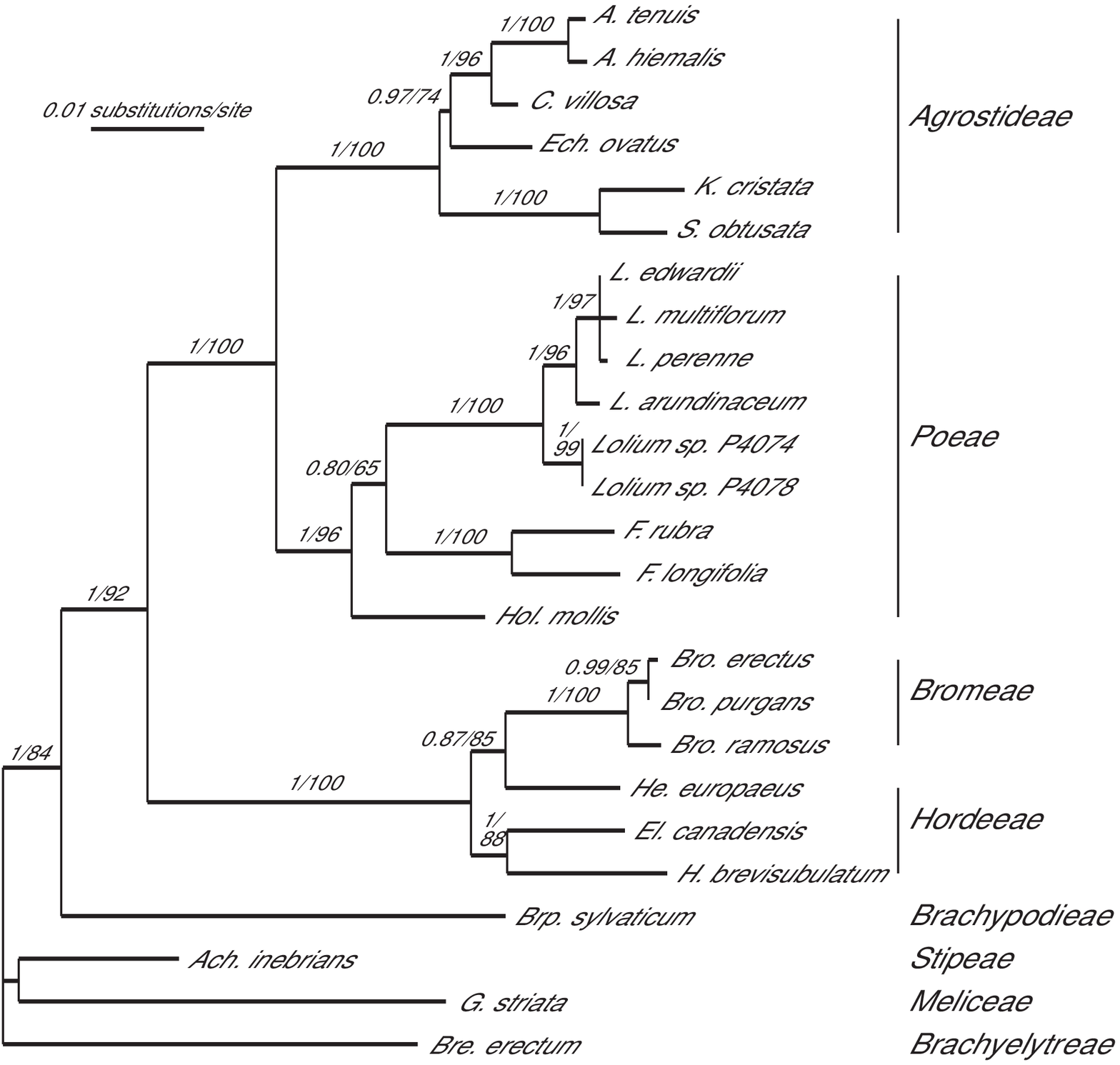}
\end{figure}

\pagebreak

{\bf Figure 3.}%\ref{figendo}}.

\begin{figure}[ht!]
\includegraphics[scale= 0.8]{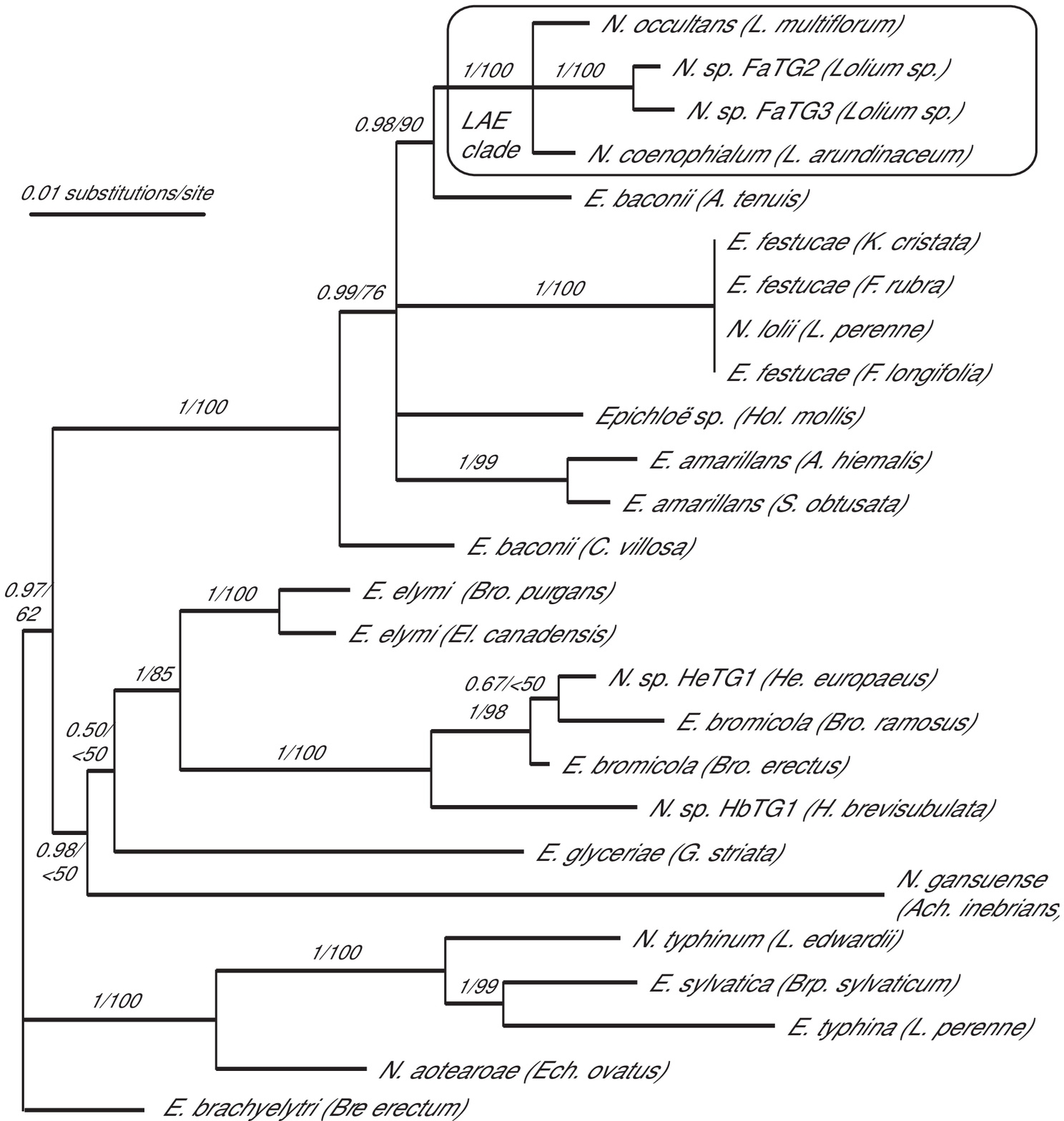}
\end{figure}

\pagebreak

{\bf Figure 4.}%\ref{fig1}}.

\begin{figure}[ht!]
\includegraphics[scale= 0.8]{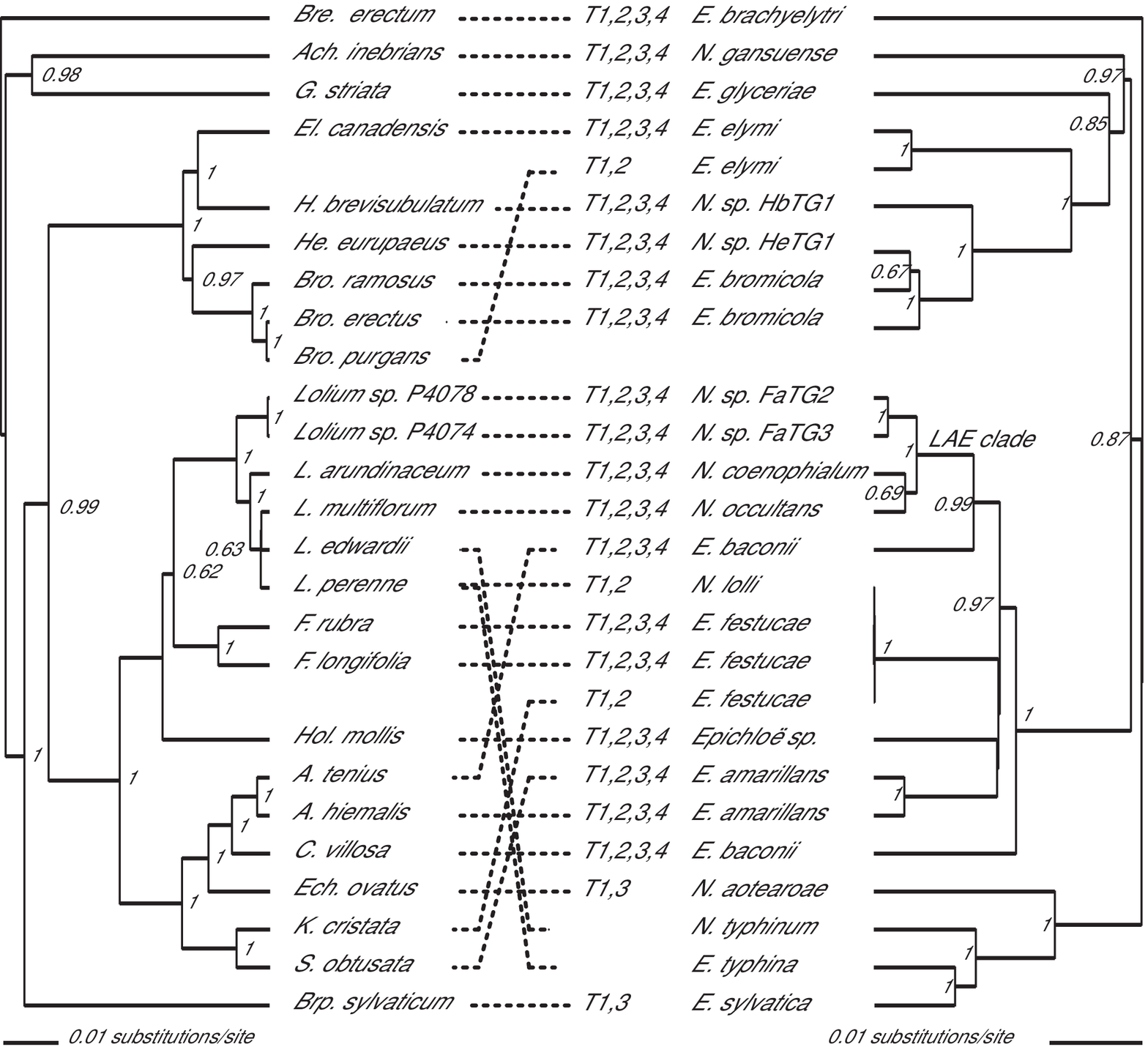}
\end{figure}

\pagebreak

{\bf Figure 5.}%\ref{fig3}}.

\begin{figure}[ht!]
\includegraphics[scale= 0.8]{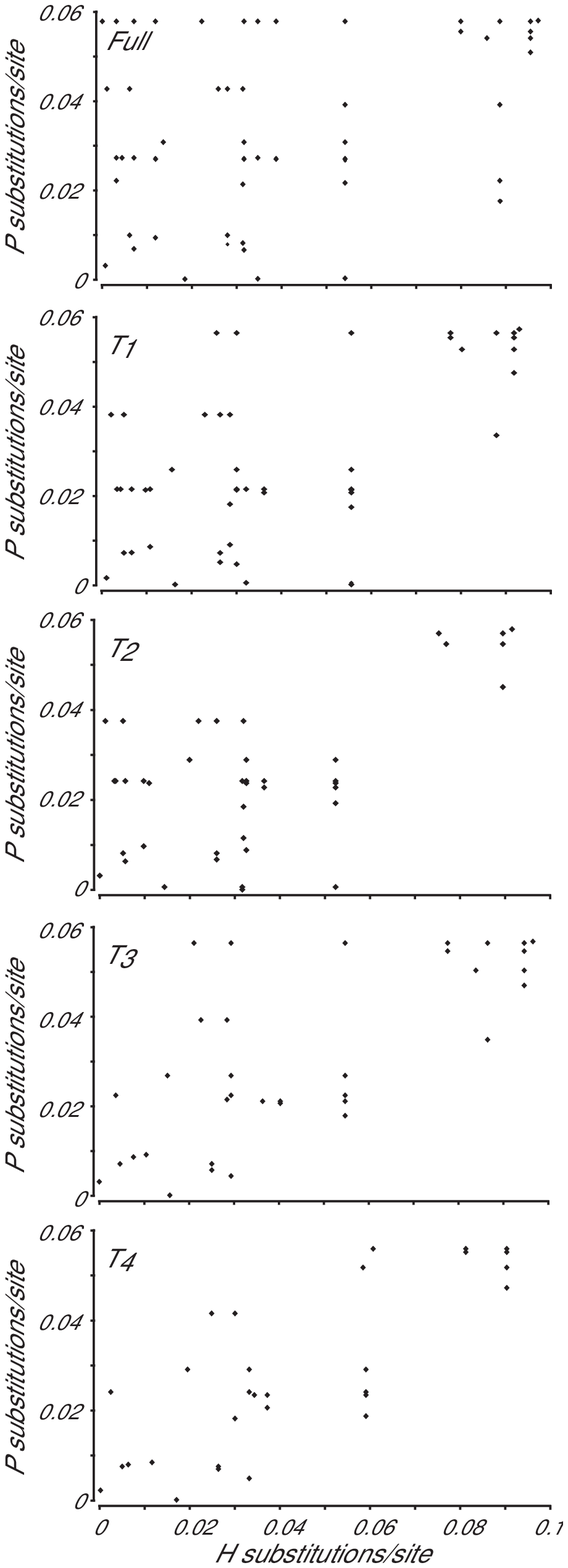}
\end{figure}

\pagebreak

{\bf Figure 6.}%\ref{GLfig}}.

\begin{figure}[ht!]
\includegraphics[scale= 1.0]{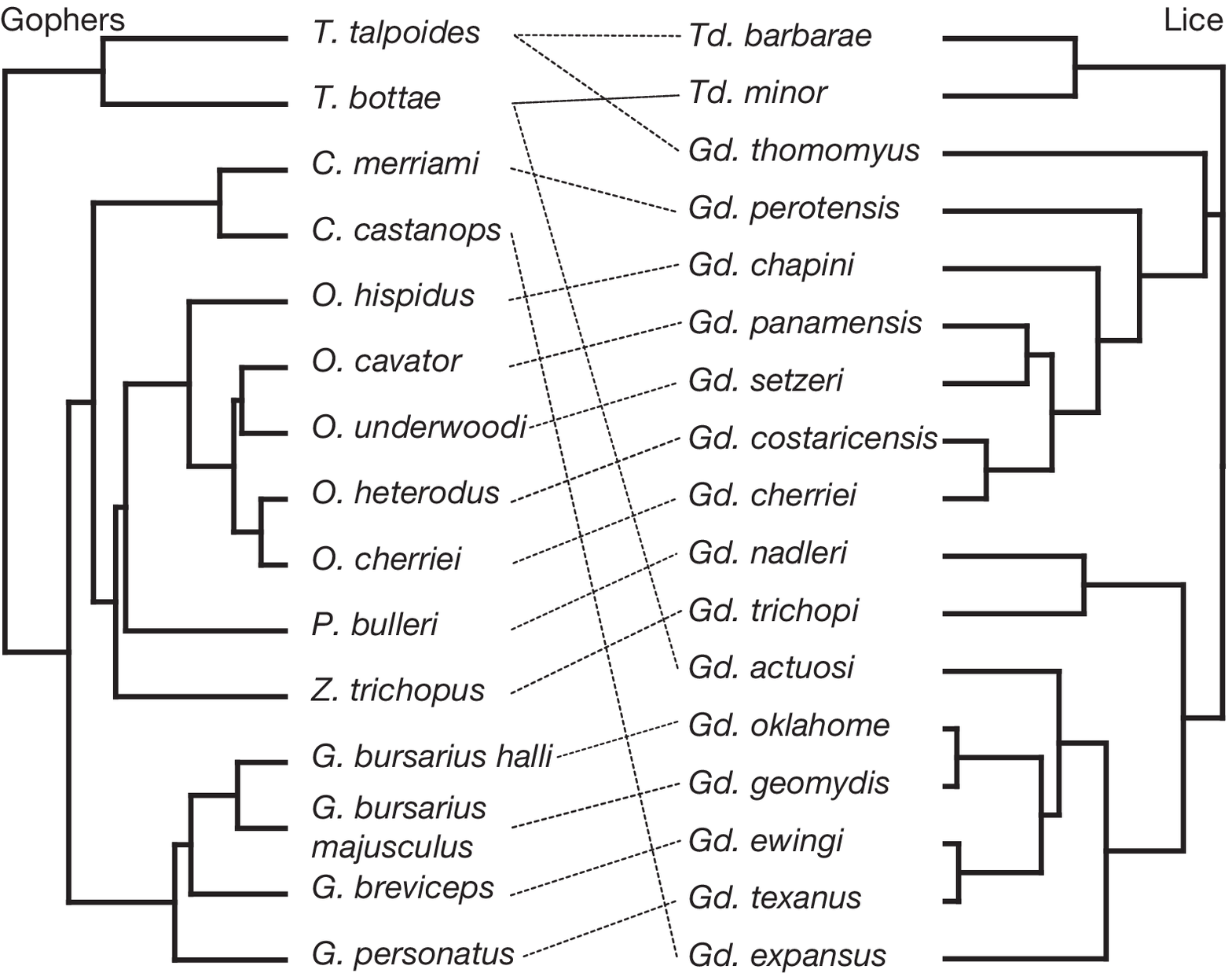}
\end{figure}


\begin{thebibliography}{sample_label}
\bibliographystyle{sysbio}
\doublespacing

\bibitem[Amenta et al.(2007)]{amenta}  Amenta, N., M. Godwin, N. Postarnakevich, and K. St. John. 2007. Approximating geodesic tree distance. Inf. Proc. Lett. 103:61--65.

\bibitem[Aris--Brosou and Yang(2003)]{Aris} Aris--Brosou, S. and Z. Yang. 2002.
Effects of models of rate evolution on estimation of divergence dates with special reference
to the metazoan 18S ribosomal RNA phylogeny. Syst. Biol. 51:703--714.

\bibitem[Billera et al.(2001)]{treespace} Billera, L. J., S. P. Holmes, and  K. Vogtmann. 2001. Geometry of the space of phylogenetic trees. Adv. Appl. Math. 27:733--767.

\bibitem[Bonnet and Peer(2002)]{mantelTest}  Bonnet, E. and Y. Van de Peer. 2002.  zt: a sofware tool for simple and partial Mantel tests. J. Stat. Softw.  7:issue 10.

\bibitem[Blanken at al.(1982)]{blanken} Blanken, R. L., L. C. Klotz, and A. G. Hinnebusch. 1982. Computer comparison of new and existing criteria for constructing evolutionary trees from sequence data.  J. Mol. Evol. 19:9--19.

\bibitem[Brem et al.(1999)]{Brem} Brem, D., and A. Leuchtmann. 1999. High prevalence of horizontal transmission of the fungal endophyte Epichlo\"e sylvatica. Bull. Geobotanical Inst. ETH 65:3--12.

\bibitem[Bull et al.(1991)]{Bull} Bull, J. J., I. J. Molineux, and W. R. Rice. 1991. Selection of benevolence in host-parasite system. Evolution 45:875--882.

\bibitem[Catal\'an et al.(2004)]{Catal} Catal\'an, P., P. Torrecilla, J. A. L. Rodriguez, and R. G. Olmstead. 2004. Phylogeny of the festucoid grasses of subtribe Loliinae and allies (Poeae, Pooideae) inferred from ITS and {\em trnL-F} sequences. Mol. Phylogenet. Evol. 31:517--541.

\bibitem[Chung and Schardl(1997)]{Chung} Chung, K.-R., and C. L. Schardl. 1997. Sexual cycle and horizontal transmission of the grass symbiont, {\em Epichlo\"e typhina}. Mycol. Res. 101:295--301.

\bibitem[Clay and Holah(1999)]{Clay99} Clay, K., and J. Holah. 1999. Fungal endophyte symbiosis and plant diversity in successional fields. Science 285:1742-1744.

\bibitem[Clay and Schardl(2002)]{Clay} Clay, K., and C. Schardl. 2002. Evolutionary origins and ecological consequences of endophyte symbiosis with grasses. Amer. Nat. 160:S99--S127.

\bibitem[Craven et al.(2001)]{Craven} Craven, K. D., P. T. W. Hsiau, A. Leuchtmann, W. Hollin, and C. L. Schardl. 2001. Multigene phylogeny of {\em Epichlo\"e} species, fungal symbionts of grasses. Ann. Missouri Bot. Gard. 88:14--34.

\bibitem[Doyle and Doyle(1990)]{Doyle} Doyle, J. J., and L. L. Doyle. 1990. Isolation of plant DNA from fresh tissue. Focus 12:13--15.

\bibitem[Drummond and Rambaut(2007)]{beast} Drummond, A. J., and A. Rambaut. 2007. BEAST: Bayesian evolutionary analysis by sampling trees. BMC Evol. Biol. 7:214.

\bibitem[Felsenstein(1981)]{Felsenstein} Felsenstein, J. 1981. Evolutionary trees from DNA sequences: a maximum likelihood approach. J. Mol. Evol. 17:368--376.

\bibitem[Freeman(1904)]{Freeman} Freeman, E. M. 1904. The seed fungus of {\em Lolium temulentum} L., the darnel. Phil. Trans. R. Soc. B 196:1--27.

\bibitem[Gentile et al.(2005)]{Gentile} Gentile, A., M. S. Rossi, D. Cabral, K. D. Craven, and C. L. Schardl. 2005. Origin, divergence, and phylogeny of epichlo\"e endophytes of native Argentine grasses. Mol. Phylogenet. Evol. 35:196--208.

\bibitem[Hafner et al.(1990)]{Hafner90} Hafner, M. S., and S. A. Nadler. 1990. Cospeciation in host parasite assemblages: comparative analysis of rates of evolution and timing of cospeciation events. Syst. Zool. 39:192-204.

\bibitem[Hafner et al.(1994)]{Hafner} Hafner, M. S., P. D. Sudman, F. X. Villablanca, T. A. Spradling, J. W. Demastes, and S. A. Nadler. 1994. Disparate rates of molecular evolution in cospeciating hosts and parasites. Science 265:1087-1090.

%\bibitem[Hafner and Page(1995)]{Hafner} Hafner, M. S., and R. D. M. Page. 1995. Molecular phylogenies and host-parasite cospeciation: gophers and lice as a model system. Phil. Trans. R. Soc. B 349:77--83.

\bibitem[Herre(1993)]{Herre} Herre, E. A. 1993. Population structure and the evolution of virulence in nematode parasites of fig wasps. Science 259:1442-1445.

\bibitem[Huelsenbeck et al.(2003)]{huelsenbeck} Huelsenbeck, J. P., B. Rannala, and B. Larget. 2003. A statistical perspective for reconstructing the history of host-parasite associations. pp. 93-119 {\em in} Tangled trees: phylogeny, cospeciation, and coevolution (R. D. M.  Page, ed.). 
University of Chicago Press, Chicago.

\bibitem[Jackson(2004)]{Jackson} Jackson, A. P. 2004. A reconciliation analysis of host switching in plant-fungal symbioses. Evolution 58:1909--1923.

\bibitem[Kellogg(2001)]{Kellogg} Kellogg, E. A. 2001. Evolutionary history of the grasses. Plant Physiol. 125:1198--1205.

\bibitem[Legendre et al.(2002)]{Legendre} Legendre, P., Y. Desdevises, and E. Bazin. 2002. A statistical test for host--parasite coevolution. Syst. Biol. 51:217--234.

\bibitem[Leuchtmann and Schardl(1998)]{Leuchtmann} Leuchtmann, A., and C. L. Schardl. 1998. Mating compatibility and phylogenetic relationships among two new species of {\em Epichlo\"e} and other congeneric European species. Mycol. Res. 102:1169--1182.

\bibitem[Moon et al.(2004)]{Moon2004} Moon, C. D., K. D. Craven, A. Leuchtmann, S. L. Clement, and C. L. Schardl. 2004. Prevalence of interspecific hybrids amongst asexual fungal endophytes of grasses. Mol. Ecol. 13:1455--1467.

\bibitem[Moon et al.(2000)]{Moon2000} Moon, C. D., B. Scott, C. L. Schardl, and M. J. Christensen. 2000. The evolutionary origins of {\em Epichlo\"e} endophytes from annual ryegrasses. Mycologia 92:1103--1118.

\bibitem[Omacini et al.(2001)]{Omacini} Omacini, M., E. J. Chaneton, C. M. Ghersa, and C. B. M\"uller. 2001. Symbiotic fungal endophytes control insect host-parasite interaction webs. Nature 409:78-81.

\bibitem[Page and Charleston(1998)]{Page} Page, R. D. M., and M. A. Charleston. 1998. Trees within trees: phylogeny and historical associations. Trends Ecol. Evol. 13:356--359.

\bibitem[Piano et al.(2005)]{Piano} Piano, E., F. B. Bertoli, M. Romani, A. Tava, L. Riccioni, M. Valvassori, A. M. Carroni, and L. Pecetti. 2005. Specificity of host-endophyte association in tall fescue populations from Sardinia, Italy. Crop Sci. 45:1456--1463.

\bibitem[Posada and Crandall(1998)]{Posada} Posada, D., and K. A. Crandall. 1998. Modeltest: testing the model of DNA substitution. Bioinformatics 14:817--818.

\bibitem[Ronquist and Huelsenbeck (2003)]{Ronquist} Ronquist, F., and J. P. Huelsenbeck. 2003. MrBayes 3: bayesian phylogenetic inference under mixed models. Bioinformatics 19:1572--1574.

\bibitem[Sampson(1937)]{Sampson37} Sampson, K. 1937. Further observations on the systemic infection of {\em Lolium}. Trans. Brit. Mycol. Soc. 21:84--97.

\bibitem[Sampson(1933)]{Sampson33} Sampson, K. 1933. The systemic infection of grasses by {\em Epichlo\"e typhina} (Pers.) Tul. Trans. Brit. Mycol. Soc. 18:30--47.

\bibitem[Sanderson(2002)]{Sanderson} Sanderson, M. J. 2002. Estimating absolute rates of molecular evolution and divergence times: A penalized likelihood approach. Mol. Biol. Evol. 19:101--109.

\bibitem[Schardl and Leuchtmann(2005)]{Schardl2005} Schardl, C. L., and A. Leuchtmann. 2005. The epichlo\"e endophytes of grasses and the symbiotic continuum. Pages 475-503 {\em in} The Fungal community: its organization and role in the ecosystem (J. Dighton, J. F. White, and P. Oudemans, eds.). CRC Press, Boca Raton, Florida.

\bibitem[Schardl and Moon(2003)]{Schardl2003} Schardl, C. L., and C. D. Moon. 2003. Processes of species evolution in Epichlo\"e/ Neotyphodium endophytes of grasses. Pages 273-310 {\em in} Clavicipitalean fungi: evolutionary biology, chemistry, biocontrol and cultural impacts (J. F. White, Jr., C. W. Bacon, N. L. Hywel-Jones, and J. W. Spatafora, eds.). Marcel-Dekker, Inc., New York and Basel.

\bibitem[Schardl et al.(1997)]{Schardl97} Schardl, C. L., A. Leuchtmann, K.-R. Chung, D. Penny, and M. R. Siegel. 1997. Coevolution by common descent of fungal symbionts ({\em Epichlo\"e} spp.) and grass hosts. Mol. Biol. Evol. 14:133--143.

\bibitem[Soreng and Davis(1998)]{Soreng} Soreng, R. J., and J. I. Davis. 1998. Phylogenetics and character evolution in the grass family (Poaceae): Simultaneous analysis of morphological and chloroplast DNA restriction site character sets. Bot. Rev. 64:1--85.

\bibitem[Sullivan and Faeth(2004)]{Sullivan} Sullivan, T. J., and S. H. Faeth. 2004. Gene flow in the endophyte {\em Neotyphodium} and implications for coevolution with {\em Festuca arizonica}. Mol. Ecol. 13:649--656.

\bibitem[Swofford(1998)]{Swofford} Swofford, D. L. 1998. PAUP*: phylogenetic analysis using parsimony (*and other methods). Sinauer Associates, Sunderland, Massachusetts.

\bibitem[Taberlet et al.(1991)]{Taberlet} Taberlet, P., L. Gielly, G. Pautou, and J. Bouvet. 1991. Universal primers for amplification of three non-coding regions of chloroplast DNA. Plant Mol. Biol. 17:1105--1109.

\bibitem[Thompson(1987)]{Thompson} Thompson, J. N. 1987. Symbiont-induced speciation. Biol. J. Linn. Soc. 32:385--393.

\bibitem[Tredway et al.(1999)]{Tredway} Tredway, L. P., J. F. White, Jr., B. S. Gaut, P. V. Reddy, M. D. Richardson, and B. B. Clarke. 1999. Phylogenetic relationships within and between {\em Epichlo\"e} and {\em Neotyphodium} endophytes as estimated by AFLP markers and rDNA sequences. Mycol. Res. 103:1593--1603.

\bibitem[Tsai et al.(1994)]{Tsai} Tsai, H. F., J. S. Liu, C. Staben, M. J. Christensen, G. Latch, M. R. Siegel, and C. L. Schardl. 1994. Evolutionary diversification of fungal endophytes of tall fescue grass by hybridization with {\em Epichlo\"e} species. Proc. Natl. Acad. Sci. USA 91:2542--2546.

\bibitem[Yang(1997)]{Yang} Yang, Z. 1997. PAML: a program package for phylogenetic analysis by maximum likelihood. CABIOS 15:555--556.

\bibitem[Yang and Rannala(1997)]{Yang2} Yang, Z. and B. Rannala. 1997. Bayesian phylogenetic inference using DNA sequences: a Markov chain Monte Carlo Method. Mol. Biol. Evol. 14:717--724.
%\end{description}
\end{thebibliography}
\end{document}